\documentclass[10pt, conference, letterpaper, nofonttune]{IEEEtran}

\usepackage[hyphens]{url}
\usepackage[utf8]{inputenc}
\usepackage{xcolor}
\usepackage{amsmath}
\usepackage{xspace}
\usepackage{epsfig}
\usepackage{balance}
\usepackage{booktabs}
\usepackage{tabularx}
\usepackage[font=footnotesize]{subcaption}
\usepackage[font=footnotesize]{caption}
\usepackage{mathtools}
\usepackage[numbers,sort&compress]{natbib}
\usepackage{soul}
\usepackage{lipsum}
\usepackage{bm}
\usepackage{amsmath,mathtools,amsfonts}

\usepackage[acronyms,nonumberlist,nopostdot,nomain,nogroupskip,acronymlists={hidden}]{glossaries}
\newglossary[algh]{hidden}{acrh}{acnh}{Hidden Acronyms}
\glsdisablehyper

\usepackage{tikz}
\usepackage{pgfplots}
\pgfplotsset{compat=newest}
\pgfplotsset{plot coordinates/math parser=false}
\newlength\fheight
\newlength\fwidth
\usetikzlibrary{plotmarks,patterns,decorations.pathreplacing,backgrounds,calc,arrows,arrows.meta,spy,matrix,scopes}
\usepgfplotslibrary{patchplots,groupplots}
\usepackage{tikzscale}
\usepackage[draft]{hyperref}
\newacronym{3gpp}{3GPP}{3rd Generation Partnership Project}
\newacronym{4g}{4G}{4th generation}
\newacronym{5g}{5G}{Fifth generation}
\newacronym{6g}{6G}{Sixth generation}
\newacronym{5gc}{5GC}{5G Core}
\newacronym{adc}{ADC}{Analog to Digital Converter}
\newacronym{aerpaw}{AERPAW}{Aerial Experimentation and Research Platform for Advanced Wireless}
\newacronym{ai}{AI}{Artificial Intelligence}
\newacronym{aimd}{AIMD}{Additive Increase Multiplicative Decrease}
\newacronym{am}{AM}{Acknowledged Mode}
\newacronym{amc}{AMC}{Adaptive Modulation and Coding}
\newacronym{amf}{AMF}{Access and Mobility Management Function}
\newacronym{aops}{AOPS}{Adaptive Order Prediction Scheduling}
\newacronym{api}{API}{Application Programming Interface}
\newacronym{apn}{APN}{Access Point Name}
\newacronym{aqm}{AQM}{Active Queue Management}
\newacronym{ausf}{AUSF}{Authentication Server Function}
\newacronym{avc}{AVC}{Advanced Video Coding}
\newacronym{awgn}{AGWN}{Additive White Gaussian Noise}
\newacronym{balia}{BALIA}{Balanced Link Adaptation Algorithm}
\newacronym{bbu}{BBU}{Base Band Unit}
\newacronym{bdp}{BDP}{Bandwidth-Delay Product}
\newacronym{ber}{BER}{Bit Error Rate}
\newacronym{bf}{BF}{Beamforming}
\newacronym{bler}{BLER}{Block Error Rate}
\newacronym{brr}{BRR}{Bayesian Ridge Regressor}
\newacronym{bsr}{BSR}{Buffer Status Report}
\newacronym{bs}{BS}{Base Station}
\newacronym{bpsk}{BPSK}{Binary Phase-shift keying}
\newacronym{bss}{BSS}{Business Support System}
\newacronym{ca}{CA}{Carrier Aggregation}
\newacronym{caas}{CaaS}{Connectivity-as-a-Service}
\newacronym{cb}{CB}{Code Block}
\newacronym{cc}{CC}{Congestion Control}
\newacronym{ccid}{CCID}{Congestion Control ID}
\newacronym{cco}{CC}{Carrier Component}
\newacronym{cdd}{CDD}{Cyclic Delay Diversity}
\newacronym{cdf}{CDF}{Cumulative Distribution Function}
\newacronym{cdn}{CDN}{Content Distribution Network}
\newacronym{cir}{CIR}{Channel Impulse Response}
\newacronym{cn}{CN}{Core Network}
\newacronym{codel}{CoDel}{Controlled Delay Management}
\newacronym{comac}{COMAC}{Converged Multi-Access and Core}
\newacronym{cord}{CORD}{Central Office Re-architected as a Datacenter}
\newacronym{cornet}{CORNET}{COgnitive Radio NETwork}
\newacronym{cosmos}{COSMOS}{Cloud Enhanced Open Software Defined Mobile Wireless Testbed for City-Scale Deployment}
\newacronym{cots}{COTS}{Commercial Off-the-Shelf}
\newacronym{cp}{CP}{Control Plane}
\newacronym{cpu}{CPU}{Central Processing Unit}
\newacronym{cqi}{CQI}{Channel Quality Information}
\newacronym{cr}{CR}{Cognitive Radio}
\newacronym{cran}{CRAN}{Cloud \gls{ran}}
\newacronym{crs}{CRS}{Cell Reference Signal}
\newacronym{csi}{CSI}{Channel State Information}
\newacronym{csirs}{CSI-RS}{Channel State Information - Reference Signal}
\newacronym{cu}{CU}{Central Unit}
\newacronym{d2tcp}{D$^2$TCP}{Deadline-aware Data center TCP}
\newacronym{d3}{D$^3$}{Deadline-Driven Delivery}
\newacronym{dac}{DAC}{Digital to Analog Converter}
\newacronym{dag}{DAG}{Directed Acyclic Graph}
\newacronym{darpa}{DARPA}{Defense Advanced Research Projects Agency}
\newacronym{das}{DAS}{Distributed Antenna System}
\newacronym{dash}{DASH}{Dynamic Adaptive Streaming over HTTP}
\newacronym{dc}{DC}{Dual Connectivity}
\newacronym{dccp}{DCCP}{Datagram Congestion Control Protocol}
\newacronym{dce}{DCE}{Direct Code Execution}
\newacronym{dci}{DCI}{Downlink Control Information}
\newacronym{dcl}{DCL}{Dear Colleague Letter}
\newacronym{dctcp}{DCTCP}{Data Center TCP}
\newacronym{dl}{DL}{Downlink}
\newacronym{dmr}{DMR}{Deadline Miss Ratio}
\newacronym{dmrs}{DMRS}{DeModulation Reference Signal}
\newacronym{drlcc}{DRL-CC}{Deep Reinforcement Learning Congestion Control}
\newacronym{drs}{DRS}{Discovery Reference Signal}
\newacronym{du}{DU}{Distributed Unit}
\newacronym{e2e}{E2E}{end-to-end}
\newacronym{ecaas}{ECaaS}{Edge-Cloud-as-a-Service}
\newacronym{ecn}{ECN}{Explicit Congestion Notification}
\newacronym{edf}{EDF}{Earliest Deadline First}
\newacronym{eirp}{EIRP}{Effective Isotropic Radiated Power}
\newacronym{em}{EM}{Electro-Magnetic}
\newacronym{embb}{eMBB}{Enhanced Mobile Broadband}
\newacronym{empower}{EMPOWER}{EMpowering transatlantic PlatfOrms for advanced WirEless Research}
\newacronym{enb}{eNB}{evolved Node Base}
\newacronym{endc}{EN-DC}{E-UTRAN-\gls{nr} \gls{dc}}
\newacronym{epc}{EPC}{Evolved Packet Core}
\newacronym{eps}{EPS}{Evolved Packet System}
\newacronym{es}{ES}{Edge Server}
\newacronym{etsi}{ETSI}{European Telecommunications Standards Institute}
\newacronym[firstplural=Estimated Times of Arrival (ETAs)]{eta}{ETA}{Estimated Time of Arrival}
\newacronym{eutran}{E-UTRAN}{Evolved Universal Terrestrial Access Network}
\newacronym{faas}{FaaS}{Function-as-a-Service}
\newacronym{fapi}{FAPI}{Functional Application Platform Interface}
\newacronym{fcc}{FCC}{Federal Communications Commission}
\newacronym{fdd}{FDD}{Frequency Division Duplexing}
\newacronym{fdm}{FDM}{Frequency Division Multiplexing}
\newacronym{fdma}{FDMA}{Frequency Division Multiple Access}
\newacronym{fed4fire}{FED4FIRE+}{Federation 4 Future Internet Research and Experimentation Plus}
\newacronym{fir}{FIR}{Finite Impulse Response}
\newacronym{fit}{FIT}{Future \acrlong{iot}}
\newacronym{fpga}{FPGA}{Field Programmable Gate Array}
\newacronym{fr2}{FR2}{Frequency Range 2}
\newacronym{fs}{FS}{Fast Switching}
\newacronym{fscc}{FSCC}{Flow Sharing Congestion Control}
\newacronym{ftp}{FTP}{File Transfer Protocol}
\newacronym{fw}{FW}{Flow Window}
\newacronym{ga128}{Ga}{Golay Sequence type A}
\newacronym{ge}{GE}{Gaussian Elimination}
\newacronym{glfsr}{GLFSR}{Galois Linear Feedback Shift Register}
\newacronym{gnb}{gNB}{Next Generation Node Base}
\newacronym{gold}{Gold}{Gold}
\newacronym{gop}{GOP}{Group of Pictures}
\newacronym{gpr}{GPR}{Gaussian Process Regressor}
\newacronym{gpu}{GPU}{Graphics Processing Unit}
\newacronym{gtp}{GTP}{GPRS Tunneling Protocol}
\newacronym{gtpc}{GTP-C}{GPRS Tunnelling Protocol Control Plane}
\newacronym{gtpu}{GTP-U}{GPRS Tunnelling Protocol User Plane}
\newacronym{gtpv2c}{GTPv2-C}{\gls{gtp} v2 - Control}
\newacronym{gw}{GW}{Gateway}
\newacronym{harq}{HARQ}{Hybrid Automatic Repeat reQuest}
\newacronym{hetnet}{HetNet}{Heterogeneous Network}
\newacronym{hh}{HH}{Hard Handover}
\newacronym{hol}{HOL}{Head-of-Line}
\newacronym{hqf}{HQF}{Highest-quality-first}
\newacronym{hss}{HSS}{Home Subscription Server}
\newacronym{http}{HTTP}{HyperText Transfer Protocol}
\newacronym{ia}{IA}{Initial Access}
\newacronym{iab}{IAB}{Integrated Access and Backhaul}
\newacronym{ic}{IC}{Incident Command}
\newacronym{ietf}{IETF}{Internet Engineering Task Force}
\newacronym{ifw}{IFW}{Interference Free Window}
\newacronym{imsi}{IMSI}{International Mobile Subscriber Identity}
\newacronym{imt}{IMT}{International Mobile Telecommunication}
\newacronym{iot}{IoT}{Internet of Things}
\newacronym{ip}{IP}{Internet Protocol}
\newacronym{iq}{IQ}{In-phase and Quadrature}
\newacronym{itu}{ITU}{International Telecommunication Union}
\newacronym{kpi}{KPI}{Key Performance Indicator}
\newacronym{kvm}{KVM}{Kernel-based Virtual Machine}
\newacronym{los}{LOS}{Line-of-Sight}
\newacronym{ls}{LS}{Loosely Synchronised}
\newacronym{lsm}{LSM}{Link-to-System Mapping}
\newacronym{lstm}{LSTM}{Long Short Term Memory}
\newacronym{lte}{LTE}{Long Term Evolution}
\newacronym{lxc}{LXC}{Linux Container}
\newacronym{m2m}{M2M}{Machine to Machine}
\newacronym{mac}{MAC}{Medium Access Control}
\newacronym{manet}{MANET}{Mobile Ad Hoc Network}
\newacronym{mano}{MANO}{Management and Orchestration}
\newacronym{mc}{MC}{Multi-Connectivity}
\newacronym{mcc}{MCC}{Mobile Cloud Computing}
\newacronym{mchem}{MCHEM}{Massive Channel Emulator}
\newacronym{mcs}{MCS}{Modulation and Coding Scheme}
\newacronym{mec}{MEC}{Multi-access Edge Computing}
\newacronym{mec2}{MEC}{Mobile Edge Cloud}
\newacronym{mfc}{MFC}{Mobile Fog Computing}
\newacronym{mi}{MI}{Mutual Information}
\newacronym{mib}{MIB}{Master Information Block}
\newacronym{miesm}{MIESM}{Mutual Information Based Effective SINR}
\newacronym{mimo}{MIMO}{Multiple Input, Multiple Output}
\newacronym{mgen}{MGEN}{Multi-Generator}
\newacronym{ml}{ML}{Machine Learning}
\newacronym{mlr}{MLR}{Maximum-local-rate}
\newacronym[plural=\gls{mme}s,firstplural=Mobility Management Entities (MMEs)]{mme}{MME}{Mobility Management Entity}
\newacronym{mmtc}{mMTC}{Massive Machine-Type Communications}
\newacronym{mmwave}{mmWave}{millimeter wave}
\newacronym{mpdccp}{MP-DCCP}{Multipath Datagram Congestion Control Protocol}
\newacronym{mptcp}{MPTCP}{Multipath TCP}
\newacronym{mr}{MR}{Maximum Rate}
\newacronym{mrdc}{MR-DC}{Multi \gls{rat} \gls{dc}}
\newacronym{mse}{MSE}{Mean Square Error}
\newacronym{mss}{MSS}{Maximum Segment Size}
\newacronym{mt}{MT}{Mobile Termination}
\newacronym{mtd}{MTD}{Machine-Type Device}
\newacronym{mtu}{MTU}{Maximum Transmission Unit}
\newacronym{mumimo}{MU-MIMO}{Multi-user \gls{mimo}}
\newacronym{mvno}{MVNO}{Mobile Virtual Network Operator}
\newacronym{nalu}{NALU}{Network Abstraction Layer Unit}
\newacronym{nas}{NAS}{Network Attached Storage}
\newacronym{nbiot}{NB-IoT}{Narrow Band IoT}
\newacronym{nfv}{NFV}{Network Function Virtualization}
\newacronym{nfvi}{NFVI}{Network Function Virtualization Infrastructure}
\newacronym{nic}{NIC}{Network Interface Card}
\newacronym{nlos}{NLOS}{Non-Line-of-Sight}
\newacronym{now}{NOW}{Non Overlapping Window}
\newacronym{nrdz}{NRDZ}{National Radio Dynamic Zone}
\newacronym{nsf}{NSF}{National Science Foundation}
\newacronym{nsm}{NSM}{Network Service Mesh}
\newacronym[type=hidden]{nr}{NR}{New Radio}
\newacronym{nrf}{NRF}{Network Repository Function}
\newacronym{nsa}{NSA}{Non Stand Alone}
\newacronym{nse}{NSE}{Network Slicing Engine}
\newacronym{nssf}{NSSF}{Network Slice Selection Function}
\newacronym{ntp}{NTP}{Network Time Protocol}
\newacronym{o2i}{O2I}{Outdoor to Indoor}
\newacronym{oai}{OAI}{OpenAirInterface}
\newacronym{oaicn}{OAI-CN}{\gls{oai} \acrlong{cn}}
\newacronym{oairan}{OAI-RAN}{\acrlong{oai} \acrlong{ran}}
\newacronym{oam}{OAM}{Operations, Administration and Maintenance}
\newacronym[plural=\gls{obu}s,firstplural=Onboard Units (OBUs)]{obu}{OBU}{Onboard Unit}
\newacronym{ofdm}{OFDM}{Orthogonal Frequency Division Multiplexing}
\newacronym{olia}{OLIA}{Opportunistic Linked Increase Algorithm}
\newacronym{omec}{OMEC}{Open Mobile Evolved Core}
\newacronym{onap}{ONAP}{Open Network Automation Platform}
\newacronym{onf}{ONF}{Open Networking Foundation}
\newacronym{onos}{ONOS}{Open Networking Operating System}
\newacronym{oom}{OOM}{\gls{onap} Operations Manager}
\newacronym{opnfv}{OPNFV}{Open Platform for \gls{nfv}}
\newacronym[type=hidden]{oran}{O-RAN}{Open \gls{ran}}
\newacronym{orbit}{ORBIT}{Open-Access Research Testbed for Next-Generation Wireless Networks}
\newacronym{os}{OS}{Operating System}
\newacronym{osc}{OSC}{O-RAN Software Community}
\newacronym{osm}{OSM}{Open Street Map}
\newacronym{oss}{OSS}{Operations Support System}
\newacronym{pa}{PA}{Position-aware}
\newacronym{pase}{PASE}{Prioritization, Arbitration, and Self-adjusting Endpoints}
\newacronym{pawr}{PAWR}{Platforms for Advanced Wireless Research}
\newacronym{pbch}{PBCH}{Physical Broadcast Channel}
\newacronym{pcef}{PCEF}{Policy and Charging Enforcement Function}
\newacronym{pcfich}{PCFICH}{Physical Control Format Indicator Channel}
\newacronym{pcrf}{PCRF}{Policy and Charging Rules Function}
\newacronym{pdcch}{PDCCH}{Physical Downlink Control Channel}
\newacronym{pdcp}{PDCP}{Packet Data Convergence Protocol}
\newacronym{pdsch}{PDSCH}{Physical Downlink Shared Channel}
\newacronym{pdu}{PDU}{Packet Data Unit}
\newacronym{pdp}{PDP}{Power Delay Profile}
\newacronym{pf}{PF}{Proportional Fair}
\newacronym{pgw}{PGW}{Packet Gateway}
\newacronym{phich}{PHICH}{Physical Hybrid ARQ Indicator Channel}
\newacronym{phy}{PHY}{Physical}
\newacronym{pl}{PL}{Path Loss}
\newacronym{pmch}{PMCH}{Physical Multicast Channel}
\newacronym{pmi}{PMI}{Precoding Matrix Indicators}
\newacronym{powder}{POWDER}{Platform for Open Wireless Data-driven Experimental Research}
\newacronym{ppo}{PPO}{Proximal Policy Optimization}
\newacronym{ppp}{PPP}{Poisson Point Process}
\newacronym{prach}{PRACH}{Physical Random Access Channel}
\newacronym{prb}{PRB}{Physical Resource Block}
\newacronym{psnr}{PSNR}{Peak Signal to Noise Ratio}
\newacronym{pss}{PSS}{Primary Synchronization Signal}
\newacronym{pucch}{PUCCH}{Physical Uplink Control Channel}
\newacronym{pusch}{PUSCH}{Physical Uplink Shared Channel}
\newacronym{qam}{QAM}{Quadrature Amplitude Modulation}
\newacronym{qci}{QCI}{\gls{qos} Class Identifier}
\newacronym{qoe}{QoE}{Quality of Experience}
\newacronym{qos}{QoS}{Quality of Service}
\newacronym{qtgui}{QT-GUI}{QT Graphical User Interface}
\newacronym{quic}{QUIC}{Quick UDP Internet Connections}
\newacronym{rach}{RACH}{Random Access Channel}
\newacronym{ran}{RAN}{Radio Access Network}
\newacronym[firstplural=Radio Access Technologies (RATs)]{rat}{RAT}{Radio Access Technology}
\newacronym{rcn}{RCN}{Research Coordination Network}
\newacronym{rec}{REC}{Radio Edge Cloud}
\newacronym{red}{RED}{Random Early Detection}
\newacronym{renew}{RENEW}{Reconfigurable Eco-system for Next-generation End-to-end Wireless}
\newacronym{rf}{RF}{Radio Frequency}
\newacronym{rfc}{RFC}{Request for Comments}
\newacronym{rfr}{RFR}{Random Forest Regressor}
\newacronym{ric}{RIC}{\gls{ran} Intelligent Controller}
\newacronym{rlc}{RLC}{Radio Link Control}
\newacronym{rlf}{RLF}{Radio Link Failure}
\newacronym{rlnc}{RLNC}{Random Linear Network Coding}
\newacronym{rmse}{RMSE}{Root Mean Squared Error}
\newacronym{rnis}{RNIS}{Radio Network Information Service}
\newacronym{rr}{RR}{Round Robin}
\newacronym{rrc}{RRC}{Radio Resource Control}
\newacronym{rrm}{RRM}{Radio Resource Management}
\newacronym{rru}{RRU}{Remote Radio Unit}
\newacronym{rs}{RS}{Remote Server}
\newacronym{rsrp}{RSRP}{Reference Signal Received Power}
\newacronym{rsrq}{RSRQ}{Reference Signal Received Quality}
\newacronym{rss}{RSS}{Received Signal Strength}
\newacronym{rssi}{RSSI}{Received Signal Strength Indicator}
\newacronym{rsu}{RSU}{Road-Side Unit}
\newacronym{rtt}{RTT}{Round Trip Time}
\newacronym{ru}{RU}{Radio Unit}
\newacronym{rw}{RW}{Receive Window}
\newacronym{rx}{RX}{Receiver}
\newacronym{s1ap}{S1AP}{S1 Application Protocol}
\newacronym{sa}{SA}{standalone}
\newacronym{sack}{SACK}{Selective Acknowledgment}
\newacronym{sap}{SAP}{Service Access Point}
\newacronym{sc2}{SC2}{Spectrum Collaboration Challenge}
\newacronym{scef}{SCEF}{Service Capability Exposure Function}
\newacronym{sch}{SCH}{Secondary Cell Handover}
\newacronym{scoot}{SCOOT}{Split Cycle Offset Optimization Technique}
\newacronym{sctp}{SCTP}{Stream Control Transmission Protocol}
\newacronym{sdap}{SDAP}{Service Data Adaptation Protocol}
\newacronym{sd}{SD}{Standard Deviation}
\newacronym{sdk}{SDK}{Software Development Kit}
\newacronym{sdm}{SDM}{Space Division Multiplexing}
\newacronym{sdma}{SDMA}{Spatial Division Multiple Access}
\newacronym{sdn}{SDN}{Software-defined Networking}
\newacronym{sdr}{SDR}{Software-defined Radio}
\newacronym{seba}{SEBA}{SDN-Enabled Broadband Access}
\newacronym{sgsn}{SGSN}{Serving GPRS Support Node}
\newacronym{sgw}{SGW}{Service Gateway}
\newacronym{si}{SI}{Study Item}
\newacronym{sib}{SIB}{Secondary Information Block}
\newacronym{sinr}{SINR}{Signal to Interference plus Noise Ratio}
\newacronym{sip}{SIP}{Session Initiation Protocol}
\newacronym{siso}{SISO}{Single Input, Single Output}
\newacronym{sla}{SLA}{Service Level Agreement}
\newacronym{sm}{SM}{Saturation Mode}
\newacronym{smf}{SMF}{Session Management Function}
\newacronym{smo}{SMO}{Service Management and Orchestration}
\newacronym{sms}{SMS}{Short Message Service}
\newacronym{smsgmsc}{SMS-GMSC}{\gls{sms}-Gateway}
\newacronym{snr}{SNR}{Signal-to-Noise-Ratio}
\newacronym{son}{SON}{Self-Organizing Network}
\newacronym{sptcp}{SPTCP}{Single Path TCP}
\newacronym{srb}{SRB}{Service Radio Bearer}
\newacronym{srn}{SRN}{Standard Radio Node}
\newacronym{srs}{SRS}{Sounding Reference Signal}
\newacronym{ss}{SS}{Synchronization Signal}
\newacronym{sss}{SSS}{Secondary Synchronization Signal}
\newacronym{st}{ST}{Spanning Tree}
\newacronym{svc}{SVC}{Scalable Video Coding}
\newacronym{tb}{TB}{Transport Block}
\newacronym{tcp}{TCP}{Transmission Control Protocol}
\newacronym{tdd}{TDD}{Time Division Duplexing}
\newacronym{tdm}{TDM}{Time Division Multiplexing}
\newacronym{tdma}{TDMA}{Time Division Multiple Access}
\newacronym{tfl}{TfL}{Transport for London}
\newacronym{tfrc}{TFRC}{TCP-Friendly Rate Control}
\newacronym{tft}{TFT}{Traffic Flow Template}
\newacronym{tgen}{TGEN}{Traffic Generator}
\newacronym{tip}{TIP}{Telecom Infra Project}
\newacronym{tm}{TM}{Transparent Mode}
\newacronym{to}{TO}{Telco Operator}
\newacronym{toa}{ToA}{Time of Arrival}
\newacronym{tr}{TR}{Technical Report}
\newacronym{trp}{TRP}{Transmitter Receiver Pair}
\newacronym{ts}{TS}{Technical Specification}
\newacronym{tti}{TTI}{Transmission Time Interval}
\newacronym{ttt}{TTT}{Time-to-Trigger}
\newacronym{tx}{TX}{Transmitter}
\newacronym{uas}{UAS}{Unmanned Aerial System}
\newacronym{uav}{UAV}{Unmanned Aerial Vehicle}
\newacronym{udm}{UDM}{Unified Data Management}
\newacronym{udp}{UDP}{User Datagram Protocol}
\newacronym{udr}{UDR}{Unified Data Repository}
\newacronym{ue}{UE}{User Equipment}
\newacronym{uhd}{UHD}{\gls{usrp} Hardware Driver}
\newacronym{ul}{UL}{Uplink}
\newacronym{um}{UM}{Unacknowledged Mode}
\newacronym{uml}{UML}{Unified Modeling Language}
\newacronym{upa}{UPA}{Uniform Planar Array}
\newacronym{upf}{UPF}{User Plane Function}
\newacronym{urllc}{URLLC}{Ultra Reliable and Low Latency Communication}
\newacronym{usa}{U.S.}{United States}
\newacronym{usim}{USIM}{Universal Subscriber Identity Module}
\newacronym{usrp}{USRP}{Universal Software Radio Peripheral}
\newacronym{utc}{UTC}{Urban Traffic Control}
\newacronym{vim}{VIM}{Virtualization Infrastructure Manager}
\newacronym{vm}{VM}{Virtual Machine}
\newacronym{vnf}{VNF}{Virtual Network Function}
\newacronym{volte}{VoLTE}{Voice over \gls{lte}}
\newacronym{voltha}{VOLTHA}{Virtual OLT HArdware Abstraction}
\newacronym{vr}{VR}{Virtual Reality}
\newacronym{vran}{vRAN}{Virtualized \gls{ran}}
\newacronym{vss}{VSS}{Video Streaming Server}
\newacronym{wbf}{WBF}{Wired Bias Function}
\newacronym{wf}{WF}{Wired-first}
\newacronym{wi}{WI}{Wireless InSite}
\newacronym{wlan}{WLAN}{Wireless Local Area Network}
\newacronym{pnf}{PNF}{Physical Network Function}
\newacronym{drl}{DRL}{Deep Reinforcement Learning}
\newacronym{mtc}{MTC}{Machine-type Communications}
\newacronym{v2x}{V2X}{Vehicle-to-everything}
\newacronym{cast}{\textit{CaST}}{Channel emulation generator and Sounder Toolchain}
\newacronym{arc}{ARC}{Aerial RAN CoLab}
\newacronym{dsp}{DSP}{Digital Signal Processing}
\newacronym{ota}{OTA}{Over-the-Air}
\newacronym{bom}{BoM}{Bill of Materials}
\newacronym{frand}{FRAND}{Fair, Reasonable, And Non-Discriminatory}
\newacronym{nvipc}{NVIPC}{NVIDIA \gls{ipc}}
\newacronym{ipc}{IPC}{Inter-Process Communication}
\newacronym{uci}{UCI}{Uplink Control Indication}
\newacronym{cbrs}{CBRS}{Citizen Broadband Radio Service}
\newacronym{ptp}{PTP}{Precision Timing Protocol}
\newacronym{scf}{SCF}{Small Cell Forum}

\newif\ifexttikz
\exttikzfalse

\ifexttikz
  \usetikzlibrary{external}
\fi

\newif\ifoverleaf
\ifthenelse{\equal{\jobname}{\detokenize{output}}}
    {\overleaftrue}
    {\overleaffalse}%

\newcommand{\testbed}{X5G\xspace}

\ifexttikz
\else
\usepackage{tikzpagenodes,etoolbox}
\usetikzlibrary{calc}
\usepackage[contents={}]{background}
\AddEverypageHook{%
\ifnumequal{\thepage}{1}{%
    \tikz[remember picture,overlay]{%
        \node[draw,
        minimum width=1.03\textwidth,
        text width=1.02\textwidth,
        font=\scriptsize
        ]
        at ($(current page header area) - (0,5pt)$)
        {%
        This paper has been accepted for publication on IEEE Workshop on Next-generation Open and Programmable Radio Access Networks (NG-OPERA), 2024. This is the authors' accepted version of the article. The final version published by IEEE is: D. Villa, I. Khan, F. Kaltenberger, N. Hedberg, R. Soares da Silva, A. Kelkar, C. Dick, S. Basagni, J. M. Jornet, T. Melodia, M. Polese, and D. Koutsonikolas, ``An Open, Programmable, Multi-vendor 5G O-RAN Testbed with NVIDIA ARC and OpenAirInterface,'' \textit{Proc. of the 2nd IEEE Workshop on Next-generation Open and Programmable Radio Access Networks (NG-OPERA)}, Vancouver, BC, Canada, May 2024.
        };
        \node[draw,
        minimum width=1.03\textwidth,
        text width=1.02\textwidth,
        font=\footnotesize
        ]
        at (current page footer area)
        {%
        ©2024 IEEE. Personal use of this material is permitted. Permission from IEEE must be obtained for all other uses, in any current or future media, including reprinting/republishing this material for advertising or promotional purposes, creating new collective works, for resale or redistribution to servers or lists, or reuse of any copyrighted component of this work in other works.
        };
    }%
}{}
}
\fi

\begin{document}
\bstctlcite{BSTcontrol}




\title{An Open, Programmable, Multi-vendor 5G O-RAN Testbed with NVIDIA ARC and OpenAirInterface}

\author{\IEEEauthorblockN{
Davide Villa\IEEEauthorrefmark{1}\textsuperscript{\textsection}, 
Imran Khan\IEEEauthorrefmark{1}\textsuperscript{\textsection}, 
Florian Kaltenberger\IEEEauthorrefmark{1}\IEEEauthorrefmark{3}, 
Nicholas Hedberg\IEEEauthorrefmark{5}, 
R\'{u}ben Soares da Silva\IEEEauthorrefmark{6},\\
Anupa Kelkar\IEEEauthorrefmark{5}, 
Chris Dick\IEEEauthorrefmark{5},
Stefano Basagni\IEEEauthorrefmark{1},
Josep M. Jornet\IEEEauthorrefmark{1},
Tommaso Melodia\IEEEauthorrefmark{1},\\
Michele Polese\IEEEauthorrefmark{1}, 
and Dimitrios Koutsonikolas\IEEEauthorrefmark{1}}

\IEEEauthorblockA{\IEEEauthorrefmark{1}Institute for the Wireless Internet of Things, Northeastern University, Boston, MA, U.S.A.\\
\IEEEauthorrefmark{3}Eurecom, Sophia Antipolis, France, 
\IEEEauthorrefmark{5}NVIDIA, Inc, Santa Clara, CA, U.S.A.,
\IEEEauthorrefmark{6}Allbesmart, Castelo Branco,
Portugal}\\
\thanks{This work was partially supported by the U.S.\ National Science Foundation under grant CNS-2117814.}

}

\makeatletter
\patchcmd{\@maketitle}
  {\addvspace{0.5\baselineskip}\egroup}
  {\addvspace{-1.5\baselineskip}\egroup}
  {}
  {}
\makeatother

\IEEEoverridecommandlockouts

\maketitle
\begingroup\renewcommand\thefootnote{\textsection}
\footnotetext{Davide Villa and Imran Khan are co-primary authors.}
\endgroup

\glsunset{nr}

\begin{abstract}
The transition of fifth generation (5G) cellular systems to softwarized, programmable, and intelligent networks depends on successfully enabling public and private \acrshort{5g} deployments that are (i) fully software-driven and (ii) with a performance at par with that of traditional monolithic systems. This requires hardware acceleration to scale the \gls{phy} layer performance, end-to-end integration and testing, and careful planning of the \gls{rf} environment. In this paper, we describe how the \testbed testbed at Northeastern University has addressed these challenges through the first 8-node network deployment of the NVIDIA \gls{arc}, with the Aerial \gls{sdk} for the PHY layer, accelerated on \gls{gpu}, and through its integration with higher layers from the \gls{oai} open-source project through the \gls{scf} \gls{fapi}. We discuss software integration, the network infrastructure, and a digital twin framework for \gls{rf} planning. We then profile the performance with up to 4 \gls{cots} smartphones for each base station with iPerf and video streaming applications, measuring a cell rate higher than $500$\:Mbps in downlink and $45$\:Mbps in uplink.
\end{abstract}


\maketitle


\glsresetall
\glsunset{usrp}
\glsunset{uhd}



\section{Introduction}
\label{sec:intro}

\gls{5g} networks have introduced significant improvements in \gls{ran} performance, with hundreds of Mbps average user data rates and extremely low latency~\cite{aarayanan2022comparative}. \gls{5g} systems are also transitioning toward more open, intelligent, and programmable architectures~\cite{polese2023understanding}. The combination of the \gls{5g} capabilities as defined by the \gls{3gpp} and the openness, softwarization, and programmability brought along by the Open \gls{ran} paradigm and O-RAN ALLIANCE have the potential to transform how we deploy and manage wireless mobile networks. Such systems leverage network disaggregation, with the layer of the \gls{5g} \gls{ran} distributed across different network functions, i.e., the \gls{cu}, the \gls{du}, and the \gls{ru}; softwarization, with the protocol stack functionalities implemented in software rather than dedicated circuits; and intelligent control, with closed-loop solutions for the \gls{ran} optimization and automation~\cite{bonati2023neutran}. 

These solutions are associated with lower capital and operational expenditures, facilitated by a supply chain with increasing diversity and robustness~\cite{dellOroRAN}, which includes open-source projects~\cite{kaltenberger2020openairinterface,gomez2016srs}. Combined with higher spectrum availability, they have paved the way toward the private \gls{5g} systems, which complement public \gls{5g} networks for site-specific use cases (e.g., industrial control, events, warehouse automation, etc).

While there are significant upsides in the transition to software-driven, disaggregated, programmable systems, there are still several challenges that need to be addressed before they can deliver \glspl{kpi} aligned with traditional cellular systems.
First, deploying end-to-end cellular systems is still a complex proposition, as automation and zero-touch provisioning and configuration are still far out in the radio domain. Second, while disaggregated solutions boast a diverse vendor ecosystem, they face challenges associated with end-to-end integration and interoperability across products~\cite{5gtesting,bahl2023accelerating,tang2023ai}. Third, completely virtualized 5G deployments need to tackle the high computational complexity of the \gls{dsp} for the \gls{phy} layer, which uses about $90$\% of the available compute when run on general-purpose CPUs. Finally, how to design \gls{ai} and \gls{ml} solutions that generalize well across a multitude of cellular network scenarios remains an active area of research.
Therefore, there is a need for a concerted effort that spans multiple communities (hardware, \gls{dsp}, software, DevOps, \gls{ai}/\gls{ml}) toward the design and deployment of open, programmable, multi-vendor cellular networks and testbeds that can support private \gls{5g} use cases and requirements with production-level stability and performance. 

In this paper, we introduce \testbed, a private \gls{5g} network testbed deployed at Northeastern University in Boston, MA, and based on a combination of open-source and programmable components from the \gls{phy} layer up to the \gls{cn}. We discuss for the first time the integration of a \gls{phy} layer implementation based on \gls{gpu} acceleration (i.e., NVIDIA Aerial) with higher layers from \gls{oai}~\cite{kaltenberger2020openairinterface}. Such integration leverages the \gls{scf} \gls{fapi} for the interaction between the \gls{mac} and \gls{phy} layers. It allows the inline acceleration of demanding \gls{phy} tasks on \gls{gpu}, hardware that is well equipped with massive parallelization of \gls{dsp} operations, enabling scalability and the embedding of \gls{ai}/\gls{ml} in the \gls{ran}.
The platform, known as NVIDIA \gls{arc}, is deployed on a dedicated multi-vendor infrastructure with 8 servers for the \gls{cu} and \gls{du}, 4 \glspl{ru} that can be installed in a lab space, O-RAN 7.2 fronthaul and timing hardware, and a dedicated \gls{5g} \gls{cn}. It delivers \glspl{kpi} representative of \gls{5g} sub-6 GHz systems, with cell throughput north of $500$\:Mbps with 4 connected \glspl{ue} and a $100$\:MHz carrier. 
NVIDIA \gls{arc} and \gls{oai} are tools that can be readily used to develop \gls{5g} and beyond intelligent use cases, thanks to the combination of a performance that improves over most open-source, non-accelerated solutions while maintaining the openness and code accessibility typical of such systems.
The rest of the paper is organized as follows.
The \gls{fapi}-based integration is described in Section~\ref{sec:arc}.
Section~\ref{sec:arc-hardware} concerns the network infrastructure.
System performance is evaluated in Section~\ref{sec:experiments} via digital twinning and multiple \gls{cots} \glspl{ue} and applications.
Section~\ref{sec:conclusions} draws conclusions and future work.


\glsunset{gnb}
\section{Full-stack Programmable \glspl{gnb}\\with NVIDIA Aerial and \gls{oai}}
\label{sec:arc}

\glsreset{arc}

\begin{figure}[t]
    \centering
    \includegraphics[width=0.70\linewidth]{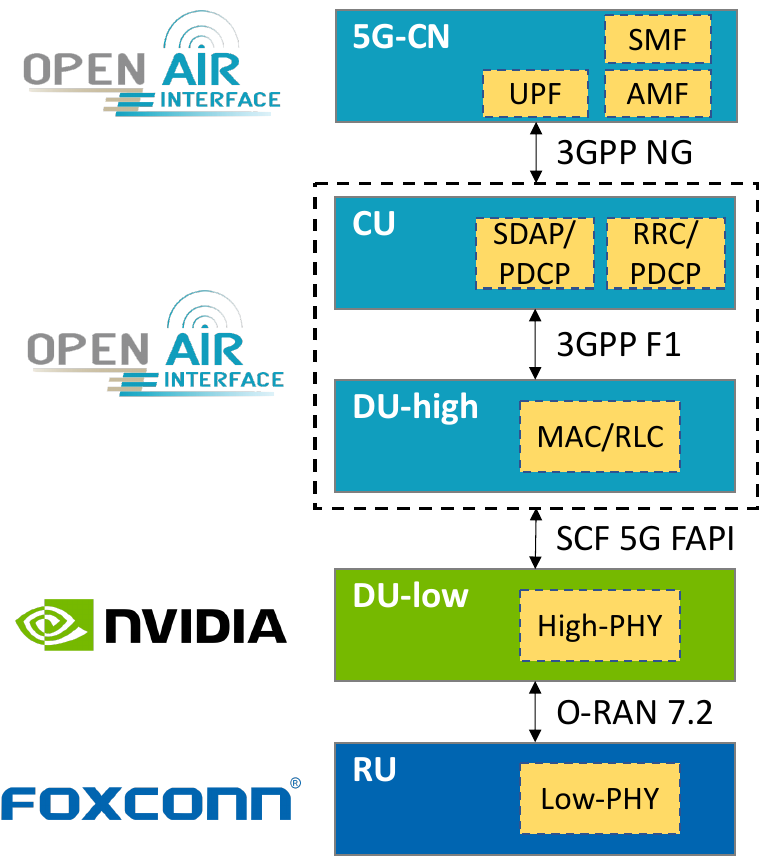}
    \caption{Architecture of the \gls{gnb} following O-RAN specifications and consisting of a Foxconn O-RU, an O-DU-low based on Nvidia Aerial \gls{sdk}, an O-DU-high and an O-CU based on \gls{oai} with their corresponding interfaces.}
    \label{fig:ARC_archi}
    \vspace{-14pt}
\end{figure}

Figure~\ref{fig:ARC_archi} shows the software architecture of the \gls{arc} \glspl{gnb}, following the basic O-RAN architecture split into \gls{cu}, \gls{du}, and \gls{ru}. The DU is further split into a DU-low, implementing Layer 1 (PHY, or L1), and into a DU-high, implementing Layer 2 (the \gls{mac} and the \gls{rlc}). They communicate over the \gls{5g} \gls{fapi} interface specified by the \gls{scf} \cite{SCF2021}. The DU-low is implemented using the NVIDIA Aerial \gls{sdk}~\cite{aerialsdk} on \gls{gpu} accelerator cards, whereas DU-high and CU are implemented by \gls{oai} on general-purpose CPUs.  In our setup, we combine the \gls{cu} and the \gls{du} to a monolithic \gls{gnb} deployed in a Docker container. Layer 1 is deployed in a separate container. 

The \gls{fapi} defines two sets of procedures. Configuration procedures handle the management of the PHY layer and happen infrequently. On the contrary, slot procedures happen in every slot (i.e. every $500$\:$\mu$s for a $30$\:kHz subcarrier spacing) and determine the structure of each slot (DL and UL). In our case, Layer 1 acts as the master and sends a slot indication every slot. Layer 2 (i.e., the \gls{mac}, or L2) replies with a UL or DL request message that specifies what the PHY needs to do at every slot. The L1 may also send indication messages to the L2 indicating reception of data for the \gls{rach}, \gls{uci}, \gls{srs}, checksums, or the user plane.  To guarantee real-time performance, the communication between the L1 and L2 containers is implemented using the \gls{nvipc} library, a generic shared memory \gls{ipc} solution capable of tracing the \gls{fapi} messages and exporting them to a pcap file analyzable with Wireshark.  

On the southbound side, the NVIDIA Layer 1 uses the O-RAN 7.2 interface \cite{ORA2023} to communicate directly with the O-RU, in our case from Foxconn. The O-RAN 7.2 transports frequency domain IQ samples with optional block floating point compression over a switched network allowing for flexible deployments. The protocol includes synchronization, control, and user planes. The S-plane is based on PTPv2. We use sync architecture option 3~\cite{oran-wg4-fronthaul-cus}, where the fronthaul switch provides synchronization to both DU and RU. The specification also includes a management plane, although it is currently not supported.

\begin{table}[b]
    \vspace{-0.2in}
    \begin{center}
    \footnotesize
    \caption{\testbed \gls{arc} deployment main features.}
    \label{table:testbeds-features}
    \begin{tabularx}{\columnwidth}{
        >{\raggedright\arraybackslash\hsize=\hsize}X
        >{\raggedright\arraybackslash\hsize=\hsize}X }
        \toprule
        Feature & Description \\
        \midrule
        \gls{3gpp} Release & 15\\
        Frequency Band & n78 (FR1, TDD)\\
        Carrier Frequency & $3.75$\:GHz\\
        Bandwidth & $100$\:MHz\\
        Subcarrier spacing & $30$\:kHz\\
        TDD config & DDDSU$^*$ \\
        Number of antennas used & 2 TX, 2 RX\\
        MIMO config & 2 layers DL, 1 layer UL \\
        Max theoretical cell throughput$^{**}$ & $525$\:Mbps DL, $94$\:Mbps UL\\
        \bottomrule
    \end{tabularx}
    \end{center}
    $^*$Currently the special slot is unused due to limitations in Foxconn radios.
    
    $^{**}$The single user maximum theoretical DL throughput is $350$\:Mbps since we can schedule a maximum of 2 DL slots per user in one TDD period, as only 2 ACK/NACK feedback bits are available per user.
\end{table}

Table~\ref{table:testbeds-features} summarizes the main features and operational parameters of the \gls{arc} deployment in the \testbed testbed. The protocol stack is aligned with \gls{3gpp} Release 15, and uses the 5G n78 \gls{tdd} band and numerology~1. The TDD pattern, which repeats every 2.5 ms, includes three downlink slots, one special slot (which is not used due to limitations in the Foxconn \glspl{ru}), and an uplink slot.

\gls{arc}, comprising the Aerial \gls{phy} and the \gls{oai} higher layers, is open and can be extended with custom features and functionalities.
The components implemented by \gls{oai} are published under the \gls{oai} public license v1.1 created by the \gls{oai} Software Alliance (OSA) in 2017~\cite{oai}.
This license is a modified Apache v2.0 License, with an additional clause that allows contributors to make patent licenses available to third parties under \gls{frand} terms, similar to \gls{3gpp} for commercial exploitation. This ensures that companies holding intellectual property in related areas can contribute. The usage of \gls{oai} code is free for non-commercial/academic research purposes. The Aerial \gls{sdk} is available through an early adopter program~\cite{aerialsdk-website}. 

\section{\testbed Architecture}
\label{sec:arc-hardware}

\begin{figure}[t]
    \centering
    \includegraphics[width=1\linewidth]{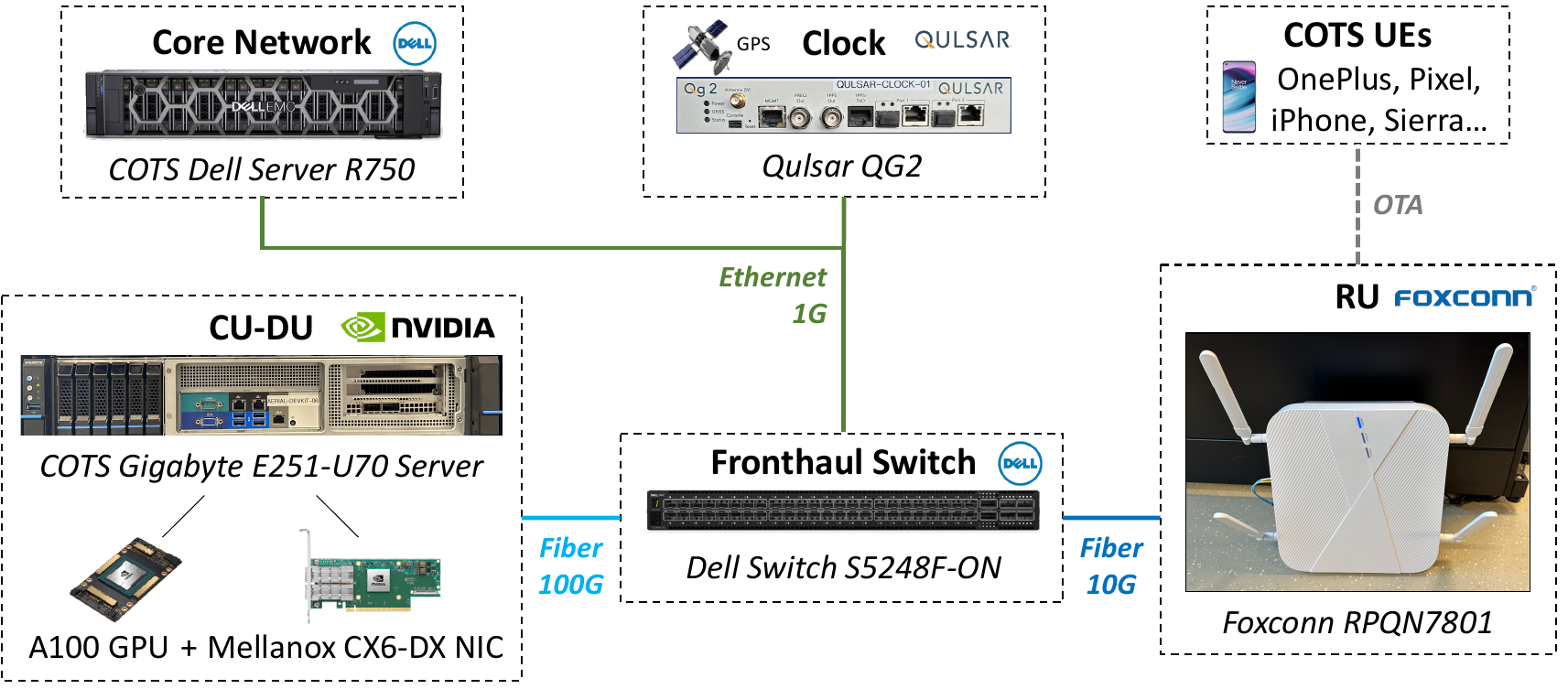}
    \caption{Hardware architecture of the \testbed deployment.}
    \label{fig:arc-hardware}
    \vspace{-10pt}
\end{figure}

This section describes the \testbed physical deployment that is currently located on the Northeastern University campus in Boston, MA\footnote{\testbed website: \url{https://x5g.org}.}. The deployment spans a server room with a dedicated rack for the private 5G system and an indoor laboratory open-space area with benches and experimental equipment that provide a realistic \gls{rf} environment with rich scattering and obstacles.
Figure~\ref{fig:arc-hardware} illustrates the hardware infrastructure that we deployed to support the \testbed operations, including 8 \gls{arc} nodes, a dedicated \gls{cn}, and a fronthaul infrastructure with timing and synchronization capabilities. 

\textbf{\gls{cu} and \gls{du}.} The 8 \gls{arc} nodes are deployed on Gigabyte E251-U70 servers, which come with (i) a half rack chassis, for deployment in \gls{ran} and edge scenarios; and (ii) a Broadcom PEX 8747 PCI switch, which enables direct connectivity between cards installed in two dedicated PCI slots, without the need for interactions with the CPU. Specifically, the two PCI slots are used for an NVIDIA A100 \gls{gpu}, which serves as the compute resource for the NVIDIA Aerial PHY layer, and for a Mellanox ConnectX-6 Dx \gls{nic}, which is used for the 7.2 fronthaul interface and equipped with two QSFP ports. Through the PCI switch, the \gls{nic} can offload or receive packets directly from the GPU, enabling low-latency packet processing. The servers are also equipped with a 24-core Intel Xeon Gold 6240R CPU and $96$\:GB of RAM.

\textbf{\gls{cn} and Backhaul.} For the \gls{cn}, we deploy the micro-services-based 5G Core from \gls{oai} on a Dell R750 server with 56 cores, $256$\:GB RAM, and multiple network interfaces. Two additional \glspl{cn} are available in the system, including Open5Gs and a commercial core from A5G, which we are integrating with \gls{arc} as part of our future work. A Dell S4112T-ON switch provides backhaul connectivity between the \glspl{gnb}, the \gls{cn} server, and the Internet, accessed through the Northeastern University switched network. 

\textbf{Fronthaul and Synchronization Infrastructure.} The fronthaul infrastructure combines switching and synchronization capabilities. It features a Dell S5248F-ON switch providing QSFP ports for the connection to the Mellanox cards on the \gls{gnb}, and 48 SFP+ ports for connectivity to the \glspl{ru}. The switch acts as a boundary clock in the synchronization plane, and it receives \gls{ptp} signals from a Qulsar QG-2 acting as grandmaster clock. The Qulsar unit is connected to a GPS antenna for precise class 6 timing, and generates both PTP and SyncE to provide frequency, phase, and time synchronization compliant with the ITU-T G.8265.1, G.8275.1, and G.8275.2 profiles. The switch is PTP-aware providing full on-path support, necessary to distribute phase synchronization.

\textbf{\gls{ru}.} Currently, we deploy 2 Foxconn RPQN 4T4R \glspl{ru}, operating in the $3.7-3.8$\:GHz band, with 2 additional units being tested in the lab. The units have 4 externally mounted antennas, each antenna with a $5$\:dBi gain, and $24$\:dBm of transmit power. The \gls{ota} transmissions are regulated as part of the Northeastern University \gls{fcc} Innovation Zone~\cite{FCC-IZ-Boston}, with an additional transmit attenuation of $20$\:dB per port to comply with transmit power limits and guarantee coexistence of multiple in-band \glspl{ru} in the same environment. As discussed in Section~\ref{sec:experiments}, we leverage two of these \glspl{ru} for the study in this paper, deployed following an \gls{rf} planning exercise. 
Plans are in place to procure and deploy \gls{cbrs} \glspl{ru} in an outdoor location to complete the 8-node deployment. 

\textbf{\gls{ue}.} Finally, we use \gls{cots} 5G \glspl{ue} from OnePlus (AC Nord 2003) and Sierra Wireless (EM9191) to connect \gls{ota} to \testbed.

\section{Experimental Analysis}
\label{sec:experiments}

We present (i) an \gls{rf} planning exercise to identify suitable locations for the \glspl{ru} deployment, through a ray-tracing-based digital twin framework of our indoor laboratory space; and (ii) a set of experiments that profile the performance of the system with varying channel conditions and different number of \glspl{ue}.

\subsection{RF Planning with Ray-tracing}
\label{sec:ray-tracing}

The purpose of the \gls{rf} planning study is to determine an optimal location for the \glspl{ru} within our indoor laboratory space in the Northeastern University ISEC building in Boston, MA.
We leverage ray-tracing in a detailed digital twin representation of the laboratory space, to achieve high fidelity between the real-world environment and the simulated one. We analyze how to deploy 2 \glspl{ru} by considering the \gls{sinr} between the \glspl{ru} and the \glspl{ue} as the objective function in the optimization problem. We restrict the optimization space by using a grid of 24 possible RU locations and 52 \gls{ue} test points and optimize through exhaustive search.

\begin{figure}[htb]
    \centering
    \begin{subfigure}[b]{\columnwidth}
        \includegraphics[width=0.99\columnwidth]{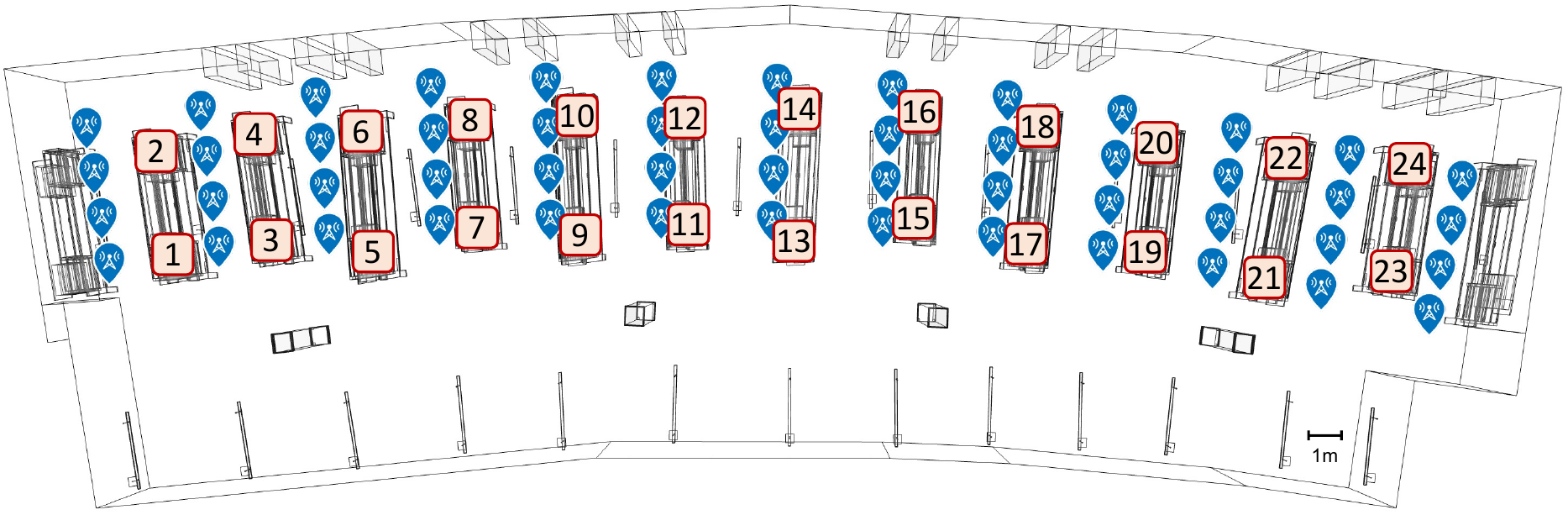}
        \caption{Site viewer top view.}
        \label{fig:siteviewertop}
    \end{subfigure} \\
    \vspace{5pt}
    \begin{subfigure}[b]{\columnwidth}
        \includegraphics[width=0.99\linewidth]{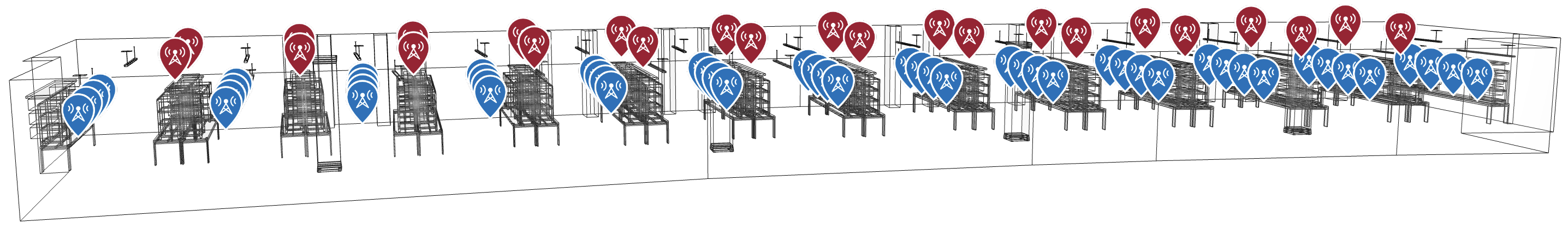}
        \caption{Site viewer side view.}
        \label{fig:siteviewerside}
    \end{subfigure}
    \caption{Site viewer with \gls{ru} (red icons) and \gls{ue} (blue icons) locations.}
    \label{fig:siteviewer}
\end{figure}

First, we leverage our digital twin framework, developed in~\cite{villa2024dt}, to create a 3D representation of our laboratory environment through the Sketchup modeling software.
We then import the model in the MATLAB ray-tracer and define the locations of \glspl{ru} and \glspl{ue} as shown in Figure~\ref{fig:siteviewertop} (from a top perspective) and in Figure~\ref{fig:siteviewerside} (from a side view). The 24 possible \glspl{ru} locations (2 for each bench) are shown in red, and the 52 test points for the \glspl{ue} are in blue. They are arranged in a $4 \times 13$ grid positioned in a semi-arc of a circle.
%
Tables~\ref{table:testbeds-features} and~\ref{table:raytracing-parameters} summarize the parameters for our ray-tracing model. In this context, we consider the \glspl{ru} as transmitter nodes (TX) transmitting to all the \glspl{ue}, which are considered only as receiver nodes (RX), i.e., we focus on optimizing the downlink rather than the uplink.
\begin{table}[h]
    \centering
    \footnotesize
    \caption{Parameters of the Matlab ray-tracing study to locate \gls{ru} locations.}
    \label{table:raytracing-parameters}
    \begin{tabularx}{0.9\columnwidth}{
        >{\raggedright\arraybackslash\hsize=1.0\hsize}X
        >{\raggedright\arraybackslash\hsize=1.0\hsize}X }
        \toprule
        Parameter & Value \\
        \midrule

        \gls{ru} antenna spacing & $0.25$\:m \\
        \gls{ru} antenna TX power ($P_{RU}$) & $24$\:dBm \\
        \gls{ru} antenna gain ($G_{RU}$) & $5$\:dBi \\
        \gls{ru} antenna pattern & Isotropic \\
        \gls{ru} TX attenuation ($A_{RU}$) & $[0-50]$\:dB \\
        Number of \gls{ru} locations & $24$ in a $2 \times 12$ grid \\
        \gls{ru} height & $2.2$\:m \\

        \gls{ue} number of antennas & 2 \\
        \gls{ue} antenna spacing & $0.07$\:m \\
        \gls{ue} antenna gain ($G_{UE}$) & $1.1$\:dBi \\
        \gls{ue} noise figure ($F_{UE}$) & $5$\:dB \\
        Number of \glspl{ue} & $52$ in a $4 \times 13$ grid \\
        \gls{ue} height & $0.8$\:m \\

        Environment material & Wood \\
        Max number of reflections & $3$ \\
        Max diffraction order & $1$ \\
        Ray-tracing method & Shooting and bouncing rays \\
        \bottomrule
    \end{tabularx}
    \vspace{-10pt}
\end{table}

The ray-tracer generates a $24 \times 52$ matrix $\mathbf{C}$ where each entry $c_{i,j}$ corresponds to the channel information between $RU_i$ and $UE_j$. From these results, we derive relevant parameters such as the thermal noise ($N$) and the path loss ($PL$) to compute 
%
the \gls{rssi} $R_{i,j}$ for $UE_j$ connected to $RU_i$, as follows:
%
%
\begin{align}\label{eq:rssi}\small
    R_{i,j} = P_{RU,i} + G_{RU,i} - A_{RU,i} - PL_{i,j} + G_{UE,i}.
\end{align}
%
Then, considering the linear representation of $\hat{R}_{i,j}$, the \gls{sinr} $\Gamma_{i,j}$ can be represented as
\begin{align}\label{eq:sinr}
    \Gamma_{i,j} = \frac{\hat{R}_{i,j}}{N F_{UE,i} + \sum\limits_{u=1, u \neq i}^M \hat{R}_{u,j}}.
\end{align}
%
%
%
Here, 
$M$ is the number of \glspl{ru} being deployed (e.g., $M=2$ in the rest of the paper).
The \gls{sinr} $\Gamma_{i,j}$ considers the interference to the signal from $RU_i$ to $UE_j$ due to downlink transmissions of all other $M - 1$ \glspl{ru} being deployed.

In this initial study, our objective is to place 2 \glspl{ru}, resulting in $M=2$, i.e., one transmits and the other creates interference.
We test all possible pairs of \glspl{ru} locations, for a total of 276 combinations. 
%
For each combination, we assign each \gls{ue} to the \gls{ru} with the best \gls{sinr} value, generating a matrix $\mathbf{\Gamma_{\max}}$ of size $276 \times 52$.
Finally, we calculate the average \gls{sinr} $\mathbb{E}(\Gamma)$ value for all \glspl{ue} in each possible deployment, a vector with 276 average \gls{sinr} values, which is used as a \emph{score} to evaluate suitable pair of locations for the \glspl{ru}.

\begin{figure}[htb]
  \centering
  \begin{subfigure}{0.32\linewidth}
    \includegraphics[width=\linewidth]{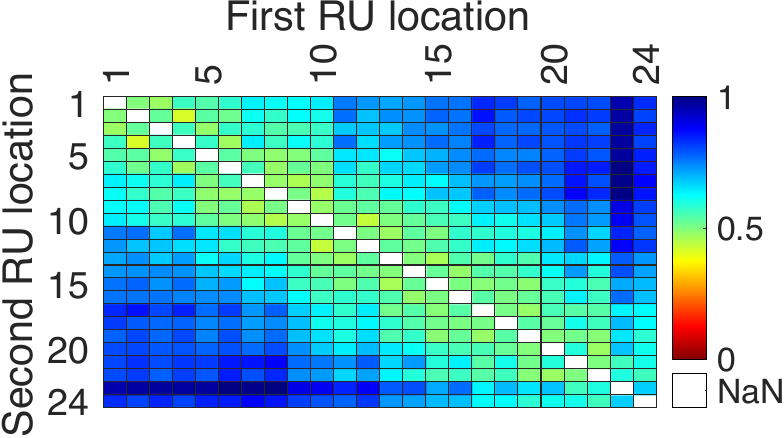}
    \caption{$0$\:dB attenuation}
    \label{fig:heatmap-scores-0db}
  \end{subfigure}
  \begin{subfigure}{0.32\linewidth}
    \includegraphics[width=\linewidth]{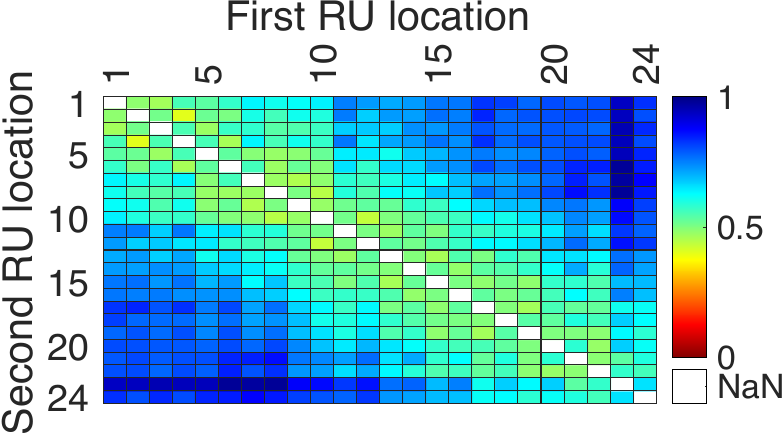}
    \caption{$10$\:dB attenuation}
    \label{fig:heatmap-scores-10db}
  \end{subfigure}
  \begin{subfigure}{0.32\linewidth}
    \includegraphics[width=\linewidth]{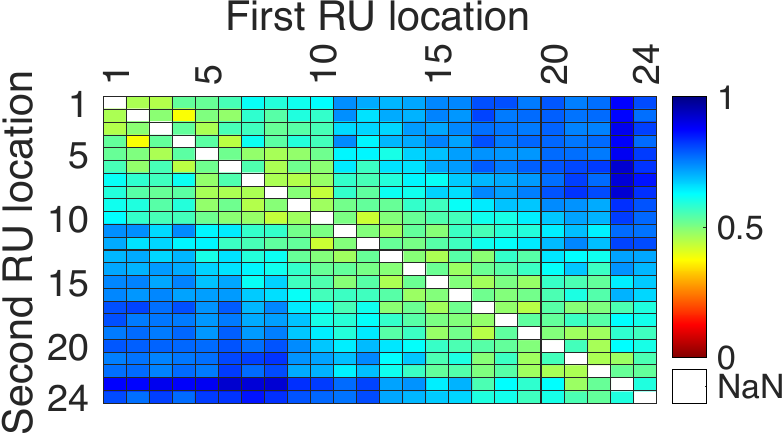}
    \caption{$20$\:dB attenuation}
    \label{fig:heatmap-scores-20db}
  \end{subfigure}
  \newline 
  \begin{subfigure}{0.32\linewidth}
    \includegraphics[width=\linewidth]{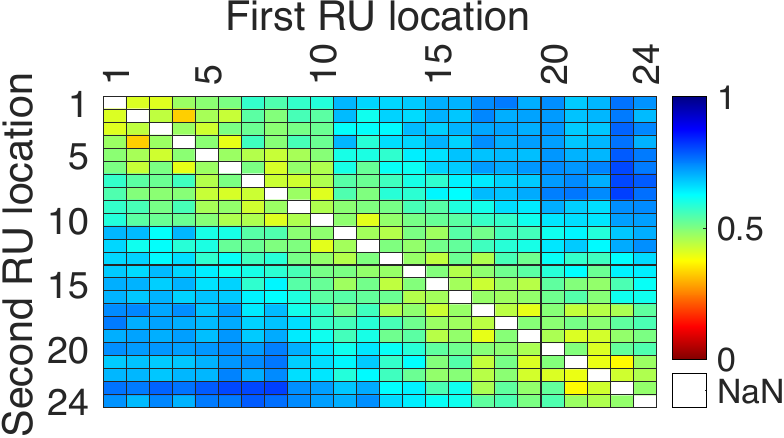}
    \caption{$30$\:dB attenuation}
    \label{fig:heatmap-scores-30db}
  \end{subfigure}
  \begin{subfigure}{0.32\linewidth}
    \includegraphics[width=\linewidth]{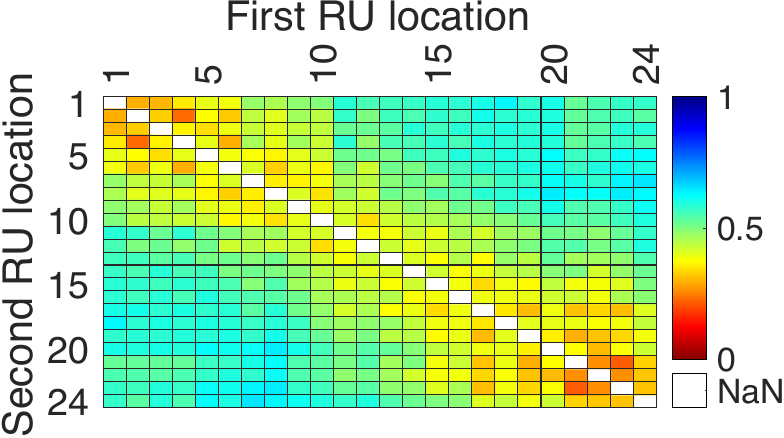}
    \caption{$40$\:dB attenuation}
    \label{fig:heatmap-scores-40db}
  \end{subfigure}
  \begin{subfigure}{0.32\linewidth}
    \includegraphics[width=\linewidth]{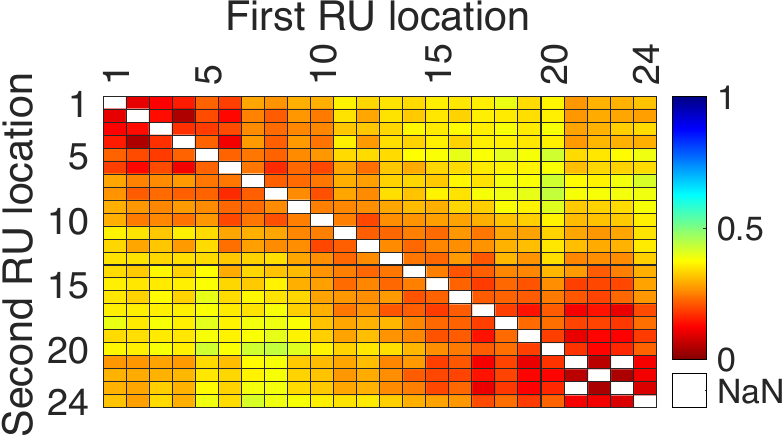}
    \caption{$50$\:dB attenuation}
    \label{fig:heatmap-scores-50db}
  \end{subfigure}
  \caption{Heatmap results of the normalized average \gls{sinr} $\mathbb{E}(\Gamma)$ with 2 \glspl{ru}.}
  \vspace{-0.1in}
  \label{fig:heatmap-scores}
\end{figure}

We test this algorithm with different values of the attenuation $A_{RU}$, from $0$ to $50$\:dB in $10$\:dB increments.
Figure~\ref{fig:heatmap-scores} visualizes the normalized values for $\mathbb{E}(\Gamma)$ for all the pairs of possible RU locations, for the different attenuation values. Additionally, Table~\ref{table:score-results} provides the best \gls{ru} locations including the minimum and maximum values of the corresponding combinations.
As expected, \glspl{ru} that are more distant exhibit high average SINR values, as they are less affected by interference. However, it is important to note that the score also considers coverage. Consequently, the optimal combination of locations identifies \glspl{ru} which are further apart but not necessarily the most distant pair.
Considering these results, for the experiments in Section~\ref{sec:exp-results}, we selected a TX attenuation of $20$\:dB which exhibits a good trade-off between coverage and the range of the average \gls{sinr} values.
Moreover, during real-world testing, we observed that a $20$\:dB attenuation leads to increased system stability and reduced degradation compared to lower attenuation values, resulting in improved overall performance, as it reduces the likelihood of saturation at the \gls{ue} antennas.
Therefore, the current locations of the \glspl{ru} in our laboratory space are [6,23].

\begin{table}[hbt]
    \centering
    \footnotesize
    \caption{Best \glspl{ru} and $\mathbb{E}(\Gamma)$ range values.}
    \label{table:score-results}
    \begin{tabularx}{0.9\columnwidth}{
        >{\raggedright\arraybackslash\hsize=0.2\hsize}X
        >{\raggedright\arraybackslash\hsize=0.35\hsize}X
        >{\raggedright\arraybackslash\hsize=0.45\hsize}X }
        \toprule
        $A_{RU}$ [dB] & \glspl{ru} with best SINR & [Min, Max] $\mathbb{E}(\Gamma)$ [dB] \\
        \midrule
        0 & [8, 23] & [6.08, 23.33] \\
        10 & [6, 23] & [5.71, 22.66] \\
        20 & [6, 23] & [5.00, 21.03] \\
        30 & [8, 23] & [3.58, 17.82] \\
        40 & [7, 24] & [0.19, 12.94] \\
        50 & [8, 20] & [-6.23, 6.63] \\
        \bottomrule
    \end{tabularx}
    \vspace{-10pt}
\end{table}

\subsection{Experiment Results}
\label{sec:exp-results}


\begin{figure}[t]
    \centering
    \includegraphics[width=1\linewidth]{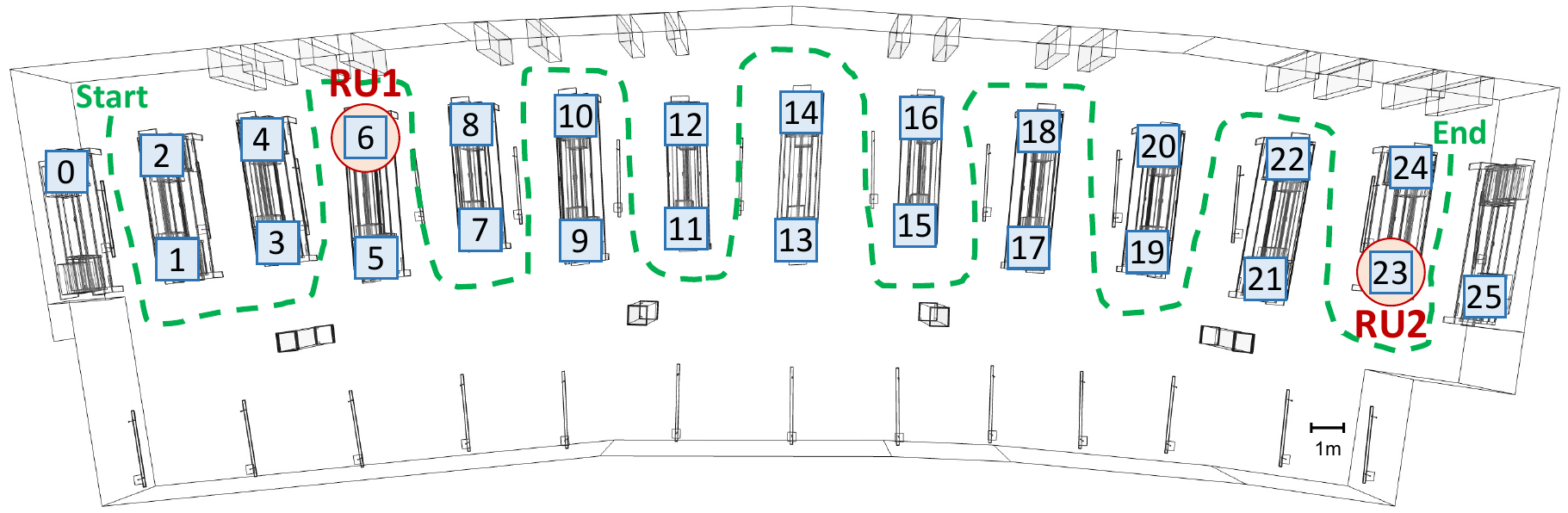}
    \vspace{-0.2in}
    \setlength\belowcaptionskip{-.5cm}
    \caption{Node locations considered in our experiments: \glspl{ru} (red circles in 6 and 23); possible static \glspl{ue} (blue squares); and mobile \glspl{ue} (green dashed line).}
    \label{fig:node-locations}
    \vspace{5pt}
\end{figure}

\begin{figure}[ht]
\centering
\begin{subfigure}[t]{0.49\textwidth}
    \centering
    \setlength\fwidth{\linewidth}
    \setlength\fheight{.4\linewidth}
\begin{tikzpicture}
\pgfplotsset{every tick label/.append style={font=\scriptsize}}

\definecolor{darkgray176}{RGB}{176,176,176}
\definecolor{darkorange25512714}{RGB}{255,127,14}
\definecolor{lightgray204}{RGB}{204,204,204}
\definecolor{steelblue31119180}{RGB}{31,119,180}

\begin{axis}[
width=0.951\fwidth,
height=\fheight,
at={(0\fwidth,0\fheight)},
x grid style={darkgray176},
xmajorticks=false,
xmin=-0.69, xmax=10.09,
xtick={0,1,2,3,4,5,6,7,8,9},
xticklabel style={rotate=45.0},
y grid style={darkgray176},
ylabel=\textcolor{steelblue31119180}{DL Throughput (Mbps)},
ylabel style={font=\scriptsize},
xlabel style={font=\scriptsize},
ymin=0, ymax=350,
ytick pos=left,
ytick style={color=steelblue31119180},
yticklabel style={color=steelblue31119180},
xmajorgrids,
ymajorgrids
]
\draw[draw=black,fill=steelblue31119180] (axis cs:-0.2,0) rectangle (axis cs:0.2,297.779273813333);

\draw[draw=black,fill=steelblue31119180] (axis cs:0.8,0) rectangle (axis cs:1.2,278.951601066667);
\draw[draw=black,fill=steelblue31119180] (axis cs:1.8,0) rectangle (axis cs:2.2,291.50394624);
\draw[draw=black,fill=steelblue31119180] (axis cs:2.8,0) rectangle (axis cs:3.2,281.087304533333);
\draw[draw=black,fill=steelblue31119180] (axis cs:3.8,0) rectangle (axis cs:4.2,286.19221888);
\draw[draw=black,fill=steelblue31119180] (axis cs:4.8,0) rectangle (axis cs:5.2,293.477207893333);
\draw[draw=black,fill=steelblue31119180] (axis cs:5.8,0) rectangle (axis cs:6.2,281.056259413333);
\draw[draw=black,fill=steelblue31119180] (axis cs:6.8,0) rectangle (axis cs:7.2,282.361967786667);
\draw[draw=black,fill=steelblue31119180] (axis cs:7.8,0) rectangle (axis cs:8.2,226.071218773333);
\draw[draw=black,fill=steelblue31119180] (axis cs:8.8,0) rectangle (axis cs:9.2,177.708098986667);
\path [draw=black, line width=1pt]
(axis cs:0,293.561317740977)
--(axis cs:0,301.997229885689);

\path [draw=black, line width=1pt]
(axis cs:1,274.913928006811)
--(axis cs:1,282.989274126523);

\path [draw=black, line width=1pt]
(axis cs:2,285.413594787131)
--(axis cs:2,297.594297692869);

\path [draw=black, line width=1pt]
(axis cs:3,274.57133522018)
--(axis cs:3,287.603273846487);

\path [draw=black, line width=1pt]
(axis cs:4,280.463374578279)
--(axis cs:4,291.921063181721);

\path [draw=black, line width=1pt]
(axis cs:5,289.103429270324)
--(axis cs:5,297.850986516343);

\path [draw=black, line width=1pt]
(axis cs:6,276.853204662378)
--(axis cs:6,285.259314164289);

\path [draw=black, line width=1pt]
(axis cs:7,278.164506604369)
--(axis cs:7,286.559428968964);

\path [draw=black, line width=1pt]
(axis cs:8,215.497162859975)
--(axis cs:8,236.645274686691);

\path [draw=black, line width=1pt]
(axis cs:9,163.498710494908)
--(axis cs:9,191.917487478425);

\addplot [semithick, black, mark=-, mark size=1.5, mark options={solid}, only marks]
table {%
0 293.561317740977
1 274.913928006811
2 285.413594787131
3 274.57133522018
4 280.463374578279
5 289.103429270324
6 276.853204662378
7 278.164506604369
8 215.497162859975
9 163.498710494908
};
\addplot [semithick, black, mark=-, mark size=1.5, mark options={solid}, only marks]
table {%
0 301.997229885689
1 282.989274126523
2 297.594297692869
3 287.603273846487
4 291.921063181721
5 297.850986516343
6 285.259314164289
7 286.559428968964
8 236.645274686691
9 191.917487478425
};
\end{axis}

\begin{axis}[
width=0.951\fwidth,
height=\fheight,
at={(0\fwidth,0\fheight)},
axis y line*=right,
legend cell align={left},
legend style={fill opacity=0.8, draw opacity=1, text opacity=1, draw=lightgray204, font=\footnotesize},
tick align=outside,
x grid style={darkgray176},
xmin=-0.69, xmax=10.09,
xtick pos=left,
ytick style={color=darkorange25512714},
xtick={0,1,2,3,4,5,6,7,8,9},
xticklabels={0,2,4,6,8,10,12,14,16,18},
y grid style={darkgray176},
ylabel=\textcolor{darkorange25512714}{UL Throughput (Mbps)},
ylabel style={font=\scriptsize},
xlabel style={font=\scriptsize},
xlabel={Location},
ymin=0, ymax=50,
ytick pos=right,
ytick style={color=darkorange25512714},
yticklabel style={anchor=west, color=darkorange25512714},
legend columns=2,
ylabel shift=-5pt
]
\draw[draw=black,fill=darkorange25512714] (axis cs:0.2,0) rectangle (axis cs:0.6,29.4579950933333);
\addlegendimage{ybar,ybar legend,draw=black,fill=darkorange25512714}
\addlegendentry{UL}
\addlegendimage{ybar,ybar legend,draw=black,fill=steelblue31119180}
\addlegendentry{DL}

\draw[draw=black,fill=darkorange25512714] (axis cs:1.2,0) rectangle (axis cs:1.6,34.2534826666667);
\draw[draw=black,fill=darkorange25512714] (axis cs:2.2,0) rectangle (axis cs:2.6,32.65265664);
\draw[draw=black,fill=darkorange25512714] (axis cs:3.2,0) rectangle (axis cs:3.6,36.9588087466667);
\draw[draw=black,fill=darkorange25512714] (axis cs:4.2,0) rectangle (axis cs:4.6,34.6589320533333);
\draw[draw=black,fill=darkorange25512714] (axis cs:5.2,0) rectangle (axis cs:5.6,25.6132164266667);
\draw[draw=black,fill=darkorange25512714] (axis cs:6.2,0) rectangle (axis cs:6.6,19.6153617066667);
\draw[draw=black,fill=darkorange25512714] (axis cs:7.2,0) rectangle (axis cs:7.6,16.84013056);
\draw[draw=black,fill=darkorange25512714] (axis cs:8.2,0) rectangle (axis cs:8.6,2.56551594666667);
\draw[draw=black,fill=darkorange25512714] (axis cs:9.2,0) rectangle (axis cs:9.6,1.18139562666667);
\path [draw=black, line width=1pt]
(axis cs:0.4,28.0220236375514)
--(axis cs:0.4,30.8939665491152);

\path [draw=black, line width=1pt]
(axis cs:1.4,33.0880112122231)
--(axis cs:1.4,35.4189541211102);

\path [draw=black, line width=1pt]
(axis cs:2.4,31.3074846346825)
--(axis cs:2.4,33.9978286453175);

\path [draw=black, line width=1pt]
(axis cs:3.4,36.5276271183574)
--(axis cs:3.4,37.3899903749759);

\path [draw=black, line width=1pt]
(axis cs:4.4,33.8209615410315)
--(axis cs:4.4,35.4969025656351);

\path [draw=black, line width=1pt]
(axis cs:5.4,24.0539830782761)
--(axis cs:5.4,27.1724497750572);

\path [draw=black, line width=1pt]
(axis cs:6.4,18.2651274938699)
--(axis cs:6.4,20.9655959194635);

\path [draw=black, line width=1pt]
(axis cs:7.4,15.5330246472272)
--(axis cs:7.4,18.1472364727728);

\path [draw=black, line width=1pt]
(axis cs:8.4,2.16213861293845)
--(axis cs:8.4,2.96889328039488);

\path [draw=black, line width=1pt]
(axis cs:9.4,0.995143241485241)
--(axis cs:9.4,1.36764801184809);

\addplot [semithick, black, mark=-, mark size=1.5, mark options={solid}, only marks]
table {%
0.4 28.0220236375514
1.4 33.0880112122231
2.4 31.3074846346825
3.4 36.5276271183574
4.4 33.8209615410315
5.4 24.0539830782761
6.4 18.2651274938699
7.4 15.5330246472272
8.4 2.16213861293845
9.4 0.995143241485241
};
\addplot [semithick, black, mark=-, mark size=1.5, mark options={solid}, only marks]
table {%
0.4 30.8939665491152
1.4 35.4189541211102
2.4 33.9978286453175
3.4 37.3899903749759
4.4 35.4969025656351
5.4 27.1724497750572
6.4 20.9655959194635
7.4 18.1472364727728
8.4 2.96889328039488
9.4 1.36764801184809
};
\end{axis}

\end{tikzpicture}
    \setlength\abovecaptionskip{-.4cm}
    \setlength\belowcaptionskip{.1cm}
    \caption{Static}
    \label{fig:static_1ru1ue}
\end{subfigure}
\begin{subfigure}[t]{0.49\textwidth}
\centering
    \setlength\fwidth{1.2\linewidth}
    \setlength\fheight{.35\linewidth}
    \input{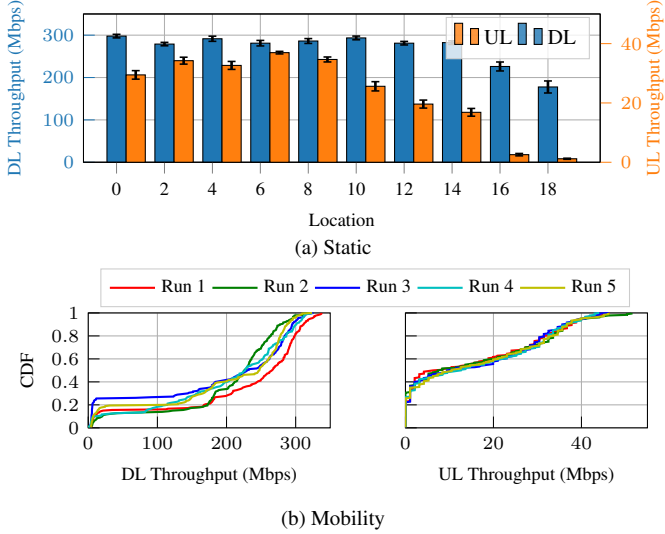}
    \caption{Mobility}
    \label{fig:mobility_1ru1ue}
\end{subfigure}
\caption{Performance profiling with one \gls{ue} and \gls{ru}, for static and mobile cases.}
\label{fig:1RU_Static}
\vspace{-20pt}
\end{figure}

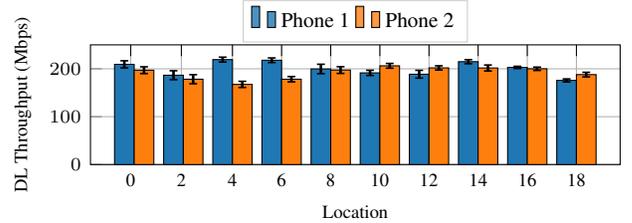
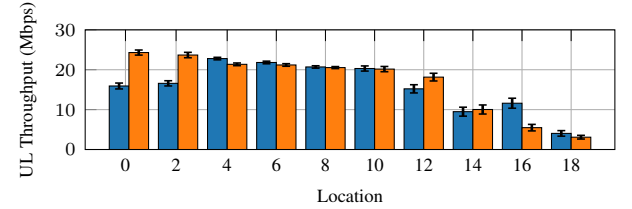
\begin{figure}[t]
\centering
\begin{subfigure}[t]{0.5\textwidth}
\centering
    \setlength\fwidth{\linewidth}
    \setlength\fheight{.35\linewidth}
\begin{tikzpicture}
\pgfplotsset{every tick label/.append style={font=\scriptsize}}

\definecolor{darkgray176}{RGB}{176,176,176}
\definecolor{darkorange25512714}{RGB}{255,127,14}
\definecolor{lightgray204}{RGB}{204,204,204}
\definecolor{steelblue31119180}{RGB}{31,119,180}

\begin{axis}[
width=0.951\fwidth,
height=\fheight,
at={(0\fwidth,0\fheight)},
legend cell align={left},
legend columns=2,
legend style={fill opacity=0.8, draw opacity=1, text opacity=1, draw=lightgray204, at={(0.5,1.02)}, anchor=south, font =\footnotesize},
x grid style={darkgray176},
xmin=-0.69, xmax=10.09,
xtick style={color=black},
xtick={0.13,1.16,2.2,3.2,4.2,5.2,6.2,7.2,8.2,9.2},
xticklabels={0,2,4,6,8,10,12,14,16,18},
y grid style={darkgray176},
ylabel={DL Throughput (Mbps)},
xlabel={Location},
ylabel style={font=\scriptsize},
xlabel style={font=\scriptsize},
ymin=0, ymax=250,
ytick pos=left,
xmajorgrids,
ymajorgrids
]
\draw[draw=black,fill=steelblue31119180] (axis cs:-0.2,0) rectangle (axis cs:0.2,209.31375104);
\addlegendimage{ybar,ybar legend,draw=black,fill=steelblue31119180}
\addlegendentry{Phone 1}

\draw[draw=black,fill=steelblue31119180] (axis cs:0.8,0) rectangle (axis cs:1.2,186.617992838095);
\draw[draw=black,fill=steelblue31119180] (axis cs:1.8,0) rectangle (axis cs:2.2,219.202990933333);
\draw[draw=black,fill=steelblue31119180] (axis cs:2.8,0) rectangle (axis cs:3.2,217.906626986667);
\draw[draw=black,fill=steelblue31119180] (axis cs:3.8,0) rectangle (axis cs:4.2,199.6595072);
\draw[draw=black,fill=steelblue31119180] (axis cs:4.8,0) rectangle (axis cs:5.2,191.50603008);
\draw[draw=black,fill=steelblue31119180] (axis cs:5.8,0) rectangle (axis cs:6.2,188.67632768);
\draw[draw=black,fill=steelblue31119180] (axis cs:6.8,0) rectangle (axis cs:7.2,214.866522026667);
\draw[draw=black,fill=steelblue31119180] (axis cs:7.8,0) rectangle (axis cs:8.2,202.916541866667);
\draw[draw=black,fill=steelblue31119180] (axis cs:8.8,0) rectangle (axis cs:9.2,175.75314304);
\draw[draw=black,fill=darkorange25512714] (axis cs:0.2,0) rectangle (axis cs:0.6,197.091720533333);
\addlegendimage{ybar,ybar legend,draw=black,fill=darkorange25512714}
\addlegendentry{Phone 2}

\draw[draw=black,fill=darkorange25512714] (axis cs:1.2,0) rectangle (axis cs:1.6,178.269318095238);
\draw[draw=black,fill=darkorange25512714] (axis cs:2.2,0) rectangle (axis cs:2.6,167.423474773333);
\draw[draw=black,fill=darkorange25512714] (axis cs:3.2,0) rectangle (axis cs:3.6,178.305252693333);
\draw[draw=black,fill=darkorange25512714] (axis cs:4.2,0) rectangle (axis cs:4.6,197.397266346667);
\draw[draw=black,fill=darkorange25512714] (axis cs:5.2,0) rectangle (axis cs:5.6,206.263659946667);
\draw[draw=black,fill=darkorange25512714] (axis cs:6.2,0) rectangle (axis cs:6.6,201.951863466667);
\draw[draw=black,fill=darkorange25512714] (axis cs:7.2,0) rectangle (axis cs:7.6,201.720223573333);
\draw[draw=black,fill=darkorange25512714] (axis cs:8.2,0) rectangle (axis cs:8.6,200.045983573333);
\draw[draw=black,fill=darkorange25512714] (axis cs:9.2,0) rectangle (axis cs:9.6,187.889971626667);
\path [draw=black, line width=1pt]
(axis cs:0,201.847342649942)
--(axis cs:0,216.780159430058);

\path [draw=black, line width=1pt]
(axis cs:1,177.31164661874)
--(axis cs:1,195.924339057451);

\path [draw=black, line width=1pt]
(axis cs:2,214.074751221685)
--(axis cs:2,224.331230644982);

\path [draw=black, line width=1pt]
(axis cs:3,213.073034039615)
--(axis cs:3,222.740219933718);

\path [draw=black, line width=1pt]
(axis cs:4,189.806104456157)
--(axis cs:4,209.512909943843);

\path [draw=black, line width=1pt]
(axis cs:5,186.048853581522)
--(axis cs:5,196.963206578478);

\path [draw=black, line width=1pt]
(axis cs:6,180.538676605231)
--(axis cs:6,196.813978754769);

\path [draw=black, line width=1pt]
(axis cs:7,210.616087254379)
--(axis cs:7,219.116956798954);

\path [draw=black, line width=1pt]
(axis cs:8,200.64366464943)
--(axis cs:8,205.189419083904);

\path [draw=black, line width=1pt]
(axis cs:9,172.587760013351)
--(axis cs:9,178.918526066649);

\addplot [semithick, black, mark=-, mark size=1.5, mark options={solid}, only marks]
table {%
0 201.847342649942
1 177.31164661874
2 214.074751221685
3 213.073034039615
4 189.806104456157
5 186.048853581522
6 180.538676605231
7 210.616087254379
8 200.64366464943
9 172.587760013351
};
\addplot [semithick, black, mark=-, mark size=1.5, mark options={solid}, only marks]
table {%
0 216.780159430058
1 195.924339057451
2 224.331230644982
3 222.740219933718
4 209.512909943843
5 196.963206578478
6 196.813978754769
7 219.116956798954
8 205.189419083904
9 178.918526066649
};
\path [draw=black, line width=1pt]
(axis cs:0.4,189.979224182537)
--(axis cs:0.4,204.20421688413);

\path [draw=black, line width=1pt]
(axis cs:1.4,168.810219824794)
--(axis cs:1.4,187.728416365682);

\path [draw=black, line width=1pt]
(axis cs:2.4,160.967087065603)
--(axis cs:2.4,173.879862481064);

\path [draw=black, line width=1pt]
(axis cs:3.4,172.736335356394)
--(axis cs:3.4,183.874170030273);

\path [draw=black, line width=1pt]
(axis cs:4.4,190.338614817949)
--(axis cs:4.4,204.455917875384);

\path [draw=black, line width=1pt]
(axis cs:5.4,201.296433481144)
--(axis cs:5.4,211.23088641219);

\path [draw=black, line width=1pt]
(axis cs:6.4,197.585972034881)
--(axis cs:6.4,206.317754898452);

\path [draw=black, line width=1pt]
(axis cs:7.4,195.473528009698)
--(axis cs:7.4,207.966919136969);

\path [draw=black, line width=1pt]
(axis cs:8.4,196.563119925935)
--(axis cs:8.4,203.528847220731);

\path [draw=black, line width=1pt]
(axis cs:9.4,183.098790508473)
--(axis cs:9.4,192.681152744861);

\addplot [semithick, black, mark=-, mark size=1.5, mark options={solid}, only marks]
table {%
0.4 189.979224182537
1.4 168.810219824794
2.4 160.967087065603
3.4 172.736335356394
4.4 190.338614817949
5.4 201.296433481144
6.4 197.585972034881
7.4 195.473528009698
8.4 196.563119925935
9.4 183.098790508473
};
\addplot [semithick, black, mark=-, mark size=1.5, mark options={solid}, only marks]
table {%
0.4 204.20421688413
1.4 187.728416365682
2.4 173.879862481064
3.4 183.874170030273
4.4 204.455917875384
5.4 211.23088641219
6.4 206.317754898452
7.4 207.966919136969
8.4 203.528847220731
9.4 192.681152744861
};
\end{axis}

\end{tikzpicture}
    \setlength\abovecaptionskip{.05cm}
    \caption{Downlink}
    \label{fig:dl_1ru2ue}
\end{subfigure}
\vspace{-0.05in}
\begin{subfigure}[t]{0.5\textwidth}
\centering
    \setlength\fwidth{\linewidth}
    \setlength\fheight{.35\linewidth}
\begin{tikzpicture}
\pgfplotsset{every tick label/.append style={font=\scriptsize}}

\definecolor{darkgray176}{RGB}{176,176,176}
\definecolor{darkorange25512714}{RGB}{255,127,14}
\definecolor{lightgray204}{RGB}{204,204,204}
\definecolor{steelblue31119180}{RGB}{31,119,180}

\begin{axis}[
width=0.951\fwidth,
height=\fheight,
at={(0\fwidth,0\fheight)},
legend cell align={left},
legend columns=2,
legend style={fill opacity=0.8, draw opacity=1, text opacity=1, draw=lightgray204, at={(0.5,1.02)}, anchor=south},
x grid style={darkgray176},
xmin=-0.69, xmax=10.09,
xtick style={color=black},
xtick={0.13,1.16,2.2,3.2,4.2,5.2,6.2,7.2,8.2,9.2},
xticklabels={0,2,4,6,8,10,12,14,16,18},
y grid style={darkgray176},
ylabel={UL Throughput (Mbps)},
xlabel={Location},
ylabel style={font=\scriptsize},
xlabel style={font=\scriptsize},
ymin=0, ymax=30,
ytick pos=left,
xmajorgrids,
ymajorgrids
]
\draw[draw=black,fill=steelblue31119180] (axis cs:-0.2,0) rectangle (axis cs:0.2,15.9243741866667);

\draw[draw=black,fill=steelblue31119180] (axis cs:0.8,0) rectangle (axis cs:1.2,16.58847232);
\draw[draw=black,fill=steelblue31119180] (axis cs:1.8,0) rectangle (axis cs:2.2,22.79604224);
\draw[draw=black,fill=steelblue31119180] (axis cs:2.8,0) rectangle (axis cs:3.2,21.8173713066667);
\draw[draw=black,fill=steelblue31119180] (axis cs:3.8,0) rectangle (axis cs:4.2,20.69889024);
\draw[draw=black,fill=steelblue31119180] (axis cs:4.8,0) rectangle (axis cs:5.2,20.3074218666667);
\draw[draw=black,fill=steelblue31119180] (axis cs:5.8,0) rectangle (axis cs:6.2,15.2113425066667);
\draw[draw=black,fill=steelblue31119180] (axis cs:6.8,0) rectangle (axis cs:7.2,9.48611754666667);
\draw[draw=black,fill=steelblue31119180] (axis cs:7.8,0) rectangle (axis cs:8.2,11.6042410666667);
\draw[draw=black,fill=steelblue31119180] (axis cs:8.8,0) rectangle (axis cs:9.2,4.04051285333333);
\draw[draw=black,fill=darkorange25512714] (axis cs:0.2,0) rectangle (axis cs:0.6,24.30599168);

\draw[draw=black,fill=darkorange25512714] (axis cs:1.2,0) rectangle (axis cs:1.6,23.6838365866667);
\draw[draw=black,fill=darkorange25512714] (axis cs:2.2,0) rectangle (axis cs:2.6,21.3559978666667);
\draw[draw=black,fill=darkorange25512714] (axis cs:3.2,0) rectangle (axis cs:3.6,21.1952162133333);
\draw[draw=black,fill=darkorange25512714] (axis cs:4.2,0) rectangle (axis cs:4.6,20.5381085866667);
\draw[draw=black,fill=darkorange25512714] (axis cs:5.2,0) rectangle (axis cs:5.6,20.1606212266667);
\draw[draw=black,fill=darkorange25512714] (axis cs:6.2,0) rectangle (axis cs:6.6,18.1473553066667);
\draw[draw=black,fill=darkorange25512714] (axis cs:7.2,0) rectangle (axis cs:7.6,10.0313770666667);
\draw[draw=black,fill=darkorange25512714] (axis cs:8.2,0) rectangle (axis cs:8.6,5.48055722666667);
\draw[draw=black,fill=darkorange25512714] (axis cs:9.2,0) rectangle (axis cs:9.6,3.08980394666667);
\path [draw=black, line width=1pt]
(axis cs:0,15.1854121927106)
--(axis cs:0,16.6633361806227);

\path [draw=black, line width=1pt]
(axis cs:1,15.9339611695867)
--(axis cs:1,17.2429834704133);

\path [draw=black, line width=1pt]
(axis cs:2,22.4809501886116)
--(axis cs:2,23.1111342913884);

\path [draw=black, line width=1pt]
(axis cs:3,21.4878175036607)
--(axis cs:3,22.1469251096727);

\path [draw=black, line width=1pt]
(axis cs:4,20.4020033048218)
--(axis cs:4,20.9957771751782);

\path [draw=black, line width=1pt]
(axis cs:5,19.6595932286659)
--(axis cs:5,20.9552505046674);

\path [draw=black, line width=1pt]
(axis cs:6,14.1801751599062)
--(axis cs:6,16.2425098534271);

\path [draw=black, line width=1pt]
(axis cs:7,8.36420179649132)
--(axis cs:7,10.608033296842);

\path [draw=black, line width=1pt]
(axis cs:8,10.3518558129289)
--(axis cs:8,12.8566263204044);

\path [draw=black, line width=1pt]
(axis cs:9,3.34300492484811)
--(axis cs:9,4.73802078181856);

\addplot [semithick, black, mark=-, mark size=1.5, mark options={solid}, only marks]
table {%
0 15.1854121927106
1 15.9339611695867
2 22.4809501886116
3 21.4878175036607
4 20.4020033048218
5 19.6595932286659
6 14.1801751599062
7 8.36420179649132
8 10.3518558129289
9 3.34300492484811
};
\addplot [semithick, black, mark=-, mark size=1.5, mark options={solid}, only marks]
table {%
0 16.6633361806227
1 17.2429834704133
2 23.1111342913884
3 22.1469251096727
4 20.9957771751782
5 20.9552505046674
6 16.2425098534271
7 10.608033296842
8 12.8566263204044
9 4.73802078181856
};
\path [draw=black, line width=1pt]
(axis cs:0.4,23.6666234045624)
--(axis cs:0.4,24.9453599554376);

\path [draw=black, line width=1pt]
(axis cs:1.4,23.0062202721671)
--(axis cs:1.4,24.3614529011663);

\path [draw=black, line width=1pt]
(axis cs:2.4,21.0196821373176)
--(axis cs:2.4,21.6923135960157);

\path [draw=black, line width=1pt]
(axis cs:3.4,20.8733946143571)
--(axis cs:3.4,21.5170378123096);

\path [draw=black, line width=1pt]
(axis cs:4.4,20.2846734632244)
--(axis cs:4.4,20.7915437101089);

\path [draw=black, line width=1pt]
(axis cs:5.4,19.5017001671464)
--(axis cs:5.4,20.8195422861869);

\path [draw=black, line width=1pt]
(axis cs:6.4,17.1807637278958)
--(axis cs:6.4,19.1139468854375);

\path [draw=black, line width=1pt]
(axis cs:7.4,8.89414285024027)
--(axis cs:7.4,11.1686112830931);

\path [draw=black, line width=1pt]
(axis cs:8.4,4.66292328647226)
--(axis cs:8.4,6.29819116686107);

\path [draw=black, line width=1pt]
(axis cs:9.4,2.62564703322214)
--(axis cs:9.4,3.55396086011119);

\addplot [semithick, black, mark=-, mark size=1.5, mark options={solid}, only marks]
table {%
0.4 23.6666234045624
1.4 23.0062202721671
2.4 21.0196821373176
3.4 20.8733946143571
4.4 20.2846734632244
5.4 19.5017001671464
6.4 17.1807637278958
7.4 8.89414285024027
8.4 4.66292328647226
9.4 2.62564703322214
};
\addplot [semithick, black, mark=-, mark size=1.5, mark options={solid}, only marks]
table {%
0.4 24.9453599554376
1.4 24.3614529011663
2.4 21.6923135960157
3.4 21.5170378123096
4.4 20.7915437101089
5.4 20.8195422861869
6.4 19.1139468854375
7.4 11.1686112830931
8.4 6.29819116686107
9.4 3.55396086011119
};
\end{axis}

\end{tikzpicture}
    \setlength\abovecaptionskip{.05cm}
    \caption{Uplink}
    \label{fig:ul_1ru2ue}
\end{subfigure}

\caption{Performance profiling for a single \gls{ru} and two \glspl{ue}.}
\label{fig:1RU2UE_Static}
\vspace{-0.25in}
\end{figure}

To evaluate \testbed, we initially assess its stability through stress tests.
We find that each \gls{gnb} with no \gls{ue} attached has an uptime of multiple days and that continuous downlink and uplink data exchange can be sustained for at least 8 hours with one or two \glspl{ue}.
%
%
%
Finally, we successfully connect and exchange traffic with 8 \glspl{ue} simultaneously to a single \gls{gnb}, using a mix of different \glspl{ue}, including smartphones and 5G modem boards.

Next, to evaluate \testbed \glspl{kpi}, we perform two different types of experiments: (i) throughput test with iPerf; and (ii) MPEG-DASH video streaming. We configure an edge server within the campus network with minimal latency ($1-2$\:ms).
For the throughput test, we transmit \gls{tcp} traffic both in downlink (DL) and uplink (UL) directions for $40$\:seconds with varying \gls{ue} numbers.
For video streaming, we configure the server with ffmpeg~\cite{FFmpeg} to advertise 5 profiles together, with the resolutions of 1080P ($250$\:Mbps, $100$\:Mbps), 720P ($50$\:Mbps, $25$\:Mbps), and 540P ($10$\:Mbps) to the \gls{ue}. We use OnePlus Nord Phone~\cite{Oneplus} as \gls{ue}, iperf3 for generating backlogged traffic, and Google's ExoPlayer~\cite{ExoPlayer} as the video client application. For all 1 \gls{ru} tests, we use the \gls{ru} at location~6 depicted in Figure~\ref{fig:node-locations}.
For each set of experiments, we do 5 runs and plot the mean and $95$\% confidence interval of the metrics.

\textbf{1 \gls{ue}, static, iPerf.} In the first set of experiments, we profile the performance of 1 \gls{ue} per 1 \gls{ru}. We select 10 static locations (0,2,4,...,18) around the lab, starting from near the \gls{ru} to far away, as depicted in Figure~\ref{fig:node-locations}. As shown in Figure~\ref{fig:static_1ru1ue}, the \gls{ue} is able to achieve an average throughput of $300$\:Mbps in DL and $38$\:Mbps in UL in the best possible locations. As expected, the average throughput decreases as we move further away from the \gls{ru}. At location~18, which is the furthest from the \gls{ru}, we observe an average throughput of approximately $177$\:Mbps in the downlink direction and $1.5$\:Mbps in the uplink direction. 

\textbf{1 \gls{ue}, mobile, iPerf.} Next, we measure throughput for $3$ minutes while walking in a zig-zag pattern around the lab, as indicated by the dashed green line in Figure~\ref{fig:node-locations}. The results are plotted in Figure~\ref{fig:mobility_1ru1ue}. It is noticeable that even during mobility, the \gls{ue} is able to achieve $300$\:Mbps in DL and $40$\:Mbps in UL. 
Additionally, we observe that when the \gls{ue} gradually moves outside the range of the \gls{ru} towards the end of the trace, the connection drops, causing $20$\% of throughput results to be 0.

\textbf{2 \glspl{ue}, static, iPerf.} We test the performance of 2 \glspl{ue} for a single \gls{ru}. We position the \glspl{ue} at the same static locations as in the previous 1 \gls{ue} static case. The results are plotted in Figure~\ref{fig:1RU2UE_Static}. We observe that, in most cases, the \glspl{ue} are able to share bandwidth fairly. The best achievable mean aggregate throughput from both \glspl{ue} is around $400$\:Mbps in DL and $44$\:Mbps in UL. This shows that the total cell throughput can be higher than the single \gls{ue} throughput. As discussed in Table~\ref{table:testbeds-features}, this is due to a limitation in the number of transport blocks that can be acked in a single slot for a \emph{single} \gls{ue}, a condition that will be relaxed in our future work. Therefore, scheduling multiple \glspl{ue} improves the resource utilization of the system.

\textbf{1 \gls{ue}, static-mobile, video streaming.} We place the \gls{ue} at three static locations at different distances from the \gls{ru}: location 8 (close); 12 (mid); and 16 (far). We run each video session for $3$ minutes and plot the mean bitrate over 5 runs, as well as the rebuffer ratio. As expected, the average bitrate decreases and the rebuffer ratio increases as the \gls{ue} moves farther from the \gls{ru}. We also observe that the \gls{ue} can achieve a 1080p resolution and a steady mean bitrate of around $180$\:Mbps in all the static cases. Note that, unlike test results achieved through backlogged traffic, the mean bitrate for video streaming is found to be lower. The video client intermittently fetches segments (causing flows to be short), which depends on parameters, e.g., video buffer and segment size. Thus, the throughput sometimes does not ramp up to the fullest during that short period and the client ABR algorithm downgrades the bitrate based on the estimate it gets. This is due to a slow \gls{mcs} selection loop in the \gls{oai} L2, which will be improved as part of our future work. Nonetheless, this shows that our setup can support up to 8K HDR videos that require $150-300$\:Mbps bitrates according to YouTube guidelines~\cite{Youtube}. During mobility, the average bitrate is found to be $120$\:Mbps, and the rebuffer ratio increases to $15$\%. This is once again because the \gls{ue} moves away from the \gls{ru}, gradually entering low-coverage regions and eventually disconnecting.

\begin{figure}[t]
\centering
\begin{subfigure}{0.45\textwidth}
    \setlength\fwidth{\linewidth}
    \setlength\fheight{.3\linewidth}
    \input{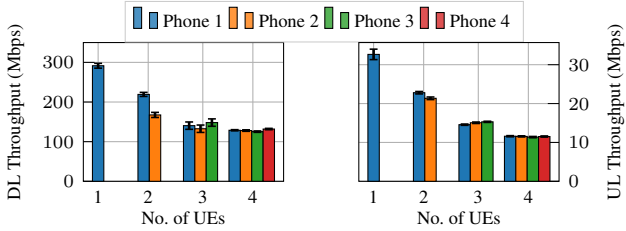}
\end{subfigure}
\vspace{-0.18in}
\caption{Performance profiling with multiple \glspl{ue}.}
\label{fig:4ue_static}
\vspace{-15pt}
\end{figure}

\textbf{1-4 \glspl{ue}, static, iPerf.}
We also extend our evaluation to multiple \gls{ue} scenarios. At a fixed location (4), we compare the performance with varying numbers of \glspl{ue} (1 to 4) connected to our network. The mean throughput and $95$\% confidence intervals are plotted in Figure~\ref{fig:4ue_static}. We observe that the \glspl{ue} are able to achieve steady throughput in all the cases. The combined throughput also increases with the increasing number of \glspl{ue} connected. With four \glspl{ue} the aggregate throughput is found to be $512$\:Mbps in DL and $46$\:Mbps in UL.

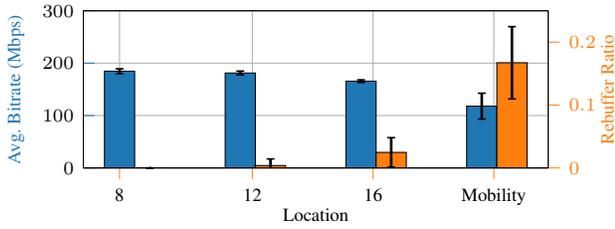
\begin{figure}[t]
\centering
\begin{subfigure}{0.45\textwidth}
    \setlength\fwidth{\linewidth}
    \setlength\fheight{.3\linewidth}
\begin{tikzpicture}
\pgfplotsset{every tick label/.append style={font=\scriptsize}}

\definecolor{darkgray176}{RGB}{176,176,176}
\definecolor{darkorange25512714}{RGB}{255,127,14}
\definecolor{lightgray204}{RGB}{204,204,204}
\definecolor{steelblue31119180}{RGB}{31,119,180}

\begin{axis}[
width=0.951\fwidth,
height=1.5\fheight,
at={(0\fwidth,0\fheight)},
x grid style={darkgray176},
xmajorticks=false,
xmin=-0.3, xmax=3.55,
xtick style={color=steelblue31119180},
xtick={0,1.1,2.1,3.1},
xticklabel style={rotate=45.0},
xticklabels={8,12,16,Mobility},
y grid style={darkgray176},
ylabel=\textcolor{steelblue31119180}{Avg. Bitrate (Mbps)},
ylabel style={font=\scriptsize},
xlabel style={font=\scriptsize},
ymin=0, ymax=300,
ytick pos=left,
ytick style={color=steelblue31119180},
xmajorgrids,
ymajorgrids
]
\draw[draw=black,fill=steelblue31119180] (axis cs:-0.125,0) rectangle (axis cs:0.125,184.684677810809);

\draw[draw=black,fill=steelblue31119180] (axis cs:0.875,0) rectangle (axis cs:1.125,181.192263115416);

\draw[draw=black,fill=steelblue31119180] (axis cs:1.875,0) rectangle (axis cs:2.125,165.600715680784);

\draw[draw=black,fill=steelblue31119180] (axis cs:2.875,0) rectangle (axis cs:3.125,117.942643215823);

\path [draw=black, line width=1pt]
(axis cs:0,180.108878846041)
--(axis cs:0,189.260476775578);

\addplot [semithick, black, mark=-, mark size=1.5, mark options={solid}, only marks]
table {%
0 180.108878846041
};
\addplot [semithick, black, mark=-, mark size=1.5, mark options={solid}, only marks]
table {%
0 189.260476775578
};
\path [draw=black, line width=1pt]
(axis cs:1,177.525601291505)
--(axis cs:1,184.858924939326);

\addplot [semithick, black, mark=-, mark size=1.5, mark options={solid}, only marks]
table {%
1 177.525601291505
};
\addplot [semithick, black, mark=-, mark size=1.5, mark options={solid}, only marks]
table {%
1 184.858924939326
};
\path [draw=black, line width=1pt]
(axis cs:2,162.956109049625)
--(axis cs:2,168.245322311942);

\addplot [semithick, black, mark=-, mark size=1.5, mark options={solid}, only marks]
table {%
2 162.956109049625
};
\addplot [semithick, black, mark=-, mark size=1.5, mark options={solid}, only marks]
table {%
2 168.245322311942
};
\path [draw=black, line width=1pt]
(axis cs:3,93.2293861692557)
--(axis cs:3,142.65590026239);

\addplot [semithick, black, mark=-, mark size=1.5, mark options={solid}, only marks]
table {%
3 93.2293861692557
};
\addplot [semithick, black, mark=-, mark size=1.5, mark options={solid}, only marks]
table {%
3 142.65590026239
};
\end{axis}

\begin{axis}[
width=0.951\fwidth,
height=1.5\fheight,
at={(0\fwidth,0\fheight)},
axis y line*=right,
legend cell align={left},
legend style={
  fill opacity=0.8,
  draw opacity=1,
  text opacity=1,
  font=\footnotesize,
  draw=lightgray204
},
tick align=outside,
x grid style={darkgray176},
xmin=-0.3, xmax=3.55,
xtick pos=left,
xtick style={color=darkorange25512714},
xtick={0,1.1,2.1,3.1},
xticklabels={8,12,16,Mobility},
x label style={at={(axis description cs:0.5,-0.20)},anchor=north},
ylabel style={font=\scriptsize},
xlabel style={font=\scriptsize},
xlabel={Location},
y grid style={darkgray176},
ylabel=\textcolor{darkorange25512714}{Rebuffer Ratio},
ymin=0, ymax=0.25,
ytick pos=right,
legend columns=4,
ytick style={color=darkorange25512714},
yticklabel style={anchor=west,color=darkorange25512714},
ylabel shift=-5pt
]
\draw[draw=black,fill=darkorange25512714] (axis cs:0.125,0) rectangle (axis cs:0.375,0);


\draw[draw=black,fill=darkorange25512714] (axis cs:1.125,0) rectangle (axis cs:1.375,0.0037973392);
\draw[draw=black,fill=darkorange25512714] (axis cs:2.125,0) rectangle (axis cs:2.375,0.0246761052);
\draw[draw=black,fill=darkorange25512714] (axis cs:3.125,0) rectangle (axis cs:3.375,0.167254825);
\path [draw=black, line width=1pt]
(axis cs:0.25,0)
--(axis cs:0.25,0);

\addplot [semithick, black, mark=-, mark size=1.5, mark options={solid}, only marks, forget plot]
table {%
0.25 0
};
\addplot [semithick, black, mark=-, mark size=1.5, mark options={solid}, only marks, forget plot]
table {%
0.25 0
};
\path [draw=black, line width=1pt]
(axis cs:1.25,-0.00674576463461573)
--(axis cs:1.25,0.0143404430346157);

\addplot [semithick, black, mark=-, mark size=1.5, mark options={solid}, only marks, forget plot]
table {%
1.25 -0.00674576463461573
};
\addplot [semithick, black, mark=-, mark size=1.5, mark options={solid}, only marks, forget plot]
table {%
1.25 0.0143404430346157
};
\path [draw=black, line width=1pt]
(axis cs:2.25,0.00120152041884096)
--(axis cs:2.25,0.048150689981159);

\addplot [semithick, black, mark=-, mark size=1.5, mark options={solid}, only marks, forget plot]
table {%
2.25 0.00120152041884096
};
\addplot [semithick, black, mark=-, mark size=1.5, mark options={solid}, only marks, forget plot]
table {%
2.25 0.048150689981159
};
\path [draw=black, line width=1pt]
(axis cs:3.25,0.1097362637768)
--(axis cs:3.25,0.2247733862232);

\addplot [semithick, black, mark=-, mark size=1.5, mark options={solid}, only marks, forget plot]
table {%
3.25 0.1097362637768
};
\addplot [semithick, black, mark=-, mark size=1.5, mark options={solid}, only marks, forget plot]
table {%
3.25 0.2247733862232
};
\end{axis}

\end{tikzpicture}
    \label{fig:static_video_1ru1ue}
\end{subfigure}
\vspace{-0.18in}
\caption{Video Streaming Performance}
\label{fig:video_static}
\vspace{-0.1in}
\end{figure}

\textbf{2 \glspl{ru}, 1 \gls{ue} per \gls{ru}, static, iPerf.} Finally, we evaluate performance with 2 \glspl{ru} by connecting phone 1 to $RU1$ and phone 2 to $RU2$. We select three pairs of locations {(0,25), (2,23), (4,21)} for the \glspl{ue} so that they are far from each other. The overall average results are listed in Table~\ref{table:2ru-case}. We observe that the impact of interference is dominant in the DL direction with each \gls{ue} experiencing up to $40$\% reduction in throughput whereas UL performance remains stable, despite a high \gls{sinr} experienced by the \glspl{ue} (average of $22.62$\:dB, reported by \gls{oai}), also aligned with the value predicted by the ray-tracer ($24.12$\:dB). We plan to analyze the impact of interference on the protocol stack in detail as part of our future work.

\begin{table}[hbt]
    \centering
    \footnotesize
    \caption{Performance profiling with 2 \glspl{ru} and 1 \gls{ue} per \gls{ru}.}
    \label{table:2ru-case}
    \begin{tabularx}{0.9\columnwidth}{
        >{\raggedright\arraybackslash\hsize=0.15\hsize}X
        >{\raggedright\arraybackslash\hsize=0.4\hsize}X
        >{\raggedright\arraybackslash\hsize=0.4\hsize}X }
        \toprule
        \gls{ue} & DL Throughput [Mbps] & UL Throughput [Mbps] \\
        \midrule
        Phone 1 & $179.32$ $\pm$ $27.29$ & $35.34$ $\pm$ $5.26$\\
        Phone 2 & $174.09$ $\pm$ $27.42$ & $33.97$ $\pm$ $2.23$ \\
        \bottomrule
    \end{tabularx}
    \vspace{-10pt}
\end{table}


\section{Conclusions and Future Works}
\label{sec:conclusions}

We introduced \testbed, an open, programmable, and multi-vendor private 5G O-RAN testbed deployed at Northeastern University in Boston, MA.
We demonstrated the first integration of NVIDIA Aerial, a PHY layer implementation on \glspl{gpu}, with higher layers based on \gls{oai}, resulting in the creation of the NVIDIA \gls{arc} platform.
We provided an overview of the \gls{arc} software and hardware implementations, designed for an 8-node deployment.
Additionally, we conducted a ray-tracing study using our digital twin framework to determine the optimal placement of \testbed \glspl{ru}, and we discussed platform performance with varying numbers of \gls{cots} \glspl{ue} and various application tests, such as iPerf and video streaming.

Next, we plan to complete the deployment of the 8 \testbed \glspl{gnb} comprising a mix of indoor and outdoor locations for more realistic experiments 
and a comprehensive development of \gls{ue} handover procedures.
We will also target the integration of \glspl{ru} supporting bands for 5G \gls{nr} \gls{fr2}.
We will develop pipelines for automatic deployment and management of workloads via container platforms like Kubernetes and Red Hat OpenShift.
Furthermore, we will perform the integration with the \gls{osc} \gls{ric}.
Our aim is to enable full control of the \gls{ran} and to ease dynamic changes of network behavior by enhancing the capabilities of \testbed for the research community of next-generation wireless networks.



\balance
\footnotesize  
\bibliographystyle{IEEEtran}
\bibliography{biblio}

\end{document}